\documentclass[aps,epsfig,amsmath,twocolumn,prb]{revtex4}
\usepackage{amssymb}
\usepackage{epsfig}
\usepackage{graphicx}
\usepackage{dcolumn}
\usepackage{bm}

\begin{document}

\newcommand{\bra}[1]{{\langle #1 |}}
\newcommand{\ket}[1]{{| #1 \rangle}}

\title{Transport through a double quantum dot in the sequential-
 and co- tunneling regimes}
\author{Vitaly N. Golovach and Daniel Loss}
\address{Department of Physics and Astronomy, University of Basel,\\
 Klingelbergstrasse 82, CH-4056 Basel, Switzerland}

\date{\today}

\begin{abstract}
We study transport through a double quantum dot, both in the sequential 
 tunneling and cotunneling regimes. 
Using a master equation approach, we find that, in the sequential tunneling 
 regime, the differential conductance 
 $G$ as a function of the bias voltage $\Delta\mu$ has a number 
 of satellite peaks with respect to the main peak of the Coulomb 
 blockade diamond. 
The position of these peaks is related to the interdot tunnel splitting and 
 the singlet-triplet splitting.
We find satellite peaks with both {\em positive} and {\em negative}
 values of differential conductance for realistic parameter regimes. 
Relating our theory to a microscopic (Hund-Mulliken) model 
 for the double dot, we find a 
 temperature regime for which the Hubbard ratio (=tunnel coupling over 
 on-site Coulomb repulsion) can be extracted from $G(\Delta\mu)$ in the
 cotunneling regime. 
In addition, we consider a combined effect of cotunneling and sequential 
 tunneling, which leads to new peaks (dips) in $G(\Delta\mu)$ inside 
 the Coulomb blockade diamond below some temperature scales, which we 
 specify.
\end{abstract}

\maketitle


                  \section{Introduction}                         %
\label{secIntro}
In recent years, there has been great interest in few-electron 
 quantum dots, or so-called artificial atoms\cite{kouwenhoven,KAT}.
The interest stems from a variety of fields, such as
 nano-electronics, spintronics, quantum computation, etc., which are all
 rapidly growing at present.
Unlike usual atoms, the quantum dots can be locally manipulated by gates 
 and tuned to the regimes of interest such that one same quantum dot 
 can realize a whole species of atomic-like electronic structures 
 (artificial atoms).
A great deal of experimental investigation on the Kondo effect, 
 Coulomb blockade effect, spin blockade effect, etc., has been performed 
 in recent years owing to this tunability feature.
Yet not a less important feature of quantum dots is the possibility to
 architecture and control, to a great extent, the coupling to their 
 surrounding, be it a dissipative  environment, classical (gate) fields, 
 or other quantum dots.
This makes quantum dots attractive for quantum computation.

A most promising candidate for qubit (quantum bit) in solid state
 is the electron spin\cite{Loss97}. 
Confining electrons to quantum dots naturally defines the qubit
 as the {\em spin up} and {\em spin down} states of the dot, 
 provided the dot
 contains an odd number of electrons. 
At the ultimate level of control over the electron charge, 
 the quantum dot can be tuned to confine one single electron\cite{note},
 thus implementing the artificial version of the hydrogen atom. 
This has been successfully achieved in recent years, 
 first in vertical dots\cite{Tarucha1} and
 lately also in lateral dots\cite{Ciorga}, 
 due to a special design of top gates. 
Furthermore, observation of shell filling of dot orbitals and
 the Hund's rule in symmetric dots\cite{Tarucha1} indicates that the electron 
 spin is a well defined and relevant degree of freedom in few-electron quantum 
 dots and that achieving control over it should be feasible experimentally 
 in the near future.

A very recent step forward in accessing the electron spin in 
 quantum dots was made in Refs.~\onlinecite{Hanson}~and~\onlinecite{Potok}, 
 where the Zeeman splitting of an electron in a lateral quantum dot has been
 measured directly by means of {\em dc} transport spectroscopy.
Resolution of the Zeeman sublevels in a magnetic field $B>5\,{\rm T}$
 allowed the authors of Ref.~\onlinecite{Hanson} 
 to measure the spin relaxation time $T_1$ 
 for a single electron by means of 
 a pulsed relaxation measurement technique.
This technique~\cite{Fujusawa} uses a sequence of pulses, which allow for 
 sequential-tunneling into excited states of the quantum dot~\cite{Fujusawa1}.
Relaxation from such an excited state can be monitored in the average 
 current versus the pulse timing, see Refs.~\onlinecite{Fujusawa} 
 and~\onlinecite{Hanson} for details.
Observation of an orbital relaxation time of $10\,{\rm ns}$~\cite{Fujusawa}
 and a $T_1$-time larger than $50\;\mu{\rm s}$ 
 at $B=7.5\,{\rm T}$~\cite{Hanson}, 
 illustrates the long-lived nature of the electron spin~\cite{noteT1}.

Integrating qubits into a quantum computer is at present at the stage 
 of realizing a two-qubit circuit. 
In the language of artificial atoms this translates to manipulating a 
 hydrogen molecule.
Here, the spins on two dots interact with the Heisenberg exchange 
 interaction $H_{\mbox{\rm spin}}=J{\bf S}_1\cdot{\bf S}_2$, where
 ${\bf S}_{1,2}$ are spin $1/2$ operators, and $J$ is the exchange 
 coupling constant.
Achieving control over the coupling constant $J$ is as important
 for the qubit as controlling its Zeeman energy for spin quantization along
 two independent axes.
The latter allows for arbitrary single-spin rotations on the Bloch sphere,
 whereas a combination of the exchange interaction and single-spin
 rotations\cite{Loss97} allows one to create non-local quantum-mechanical 
 correlations between the qubits, often referred to as {\em entanglement}.
Moreover, as demonstrated in Ref.~\onlinecite{Loss97}, the exchange 
 interaction together with single-spin rotations suffices for universal quantum
 computation.

The challenge of implementing an artificial $H_2$ molecule has 
 been met in recent works, see Refs. \onlinecite{Elzerman} 
 and \onlinecite{Pioro}.
In typical structures, transport measurements alone do not always give 
 reliable information about the number of electrons on each dot.
The current through the double dot (DD) is often too small to be measured
 and the tunnel contacts are likely to be pinched off in the
 few-electron regime, due to a high level of depletion in the dot region 
 caused by the electrostatic potential of the gates.
A complementary method of charge control was used in 
 Ref.~\onlinecite{Elzerman}, which came with integrating the quantum 
 dots with sensitive charge detectors~\cite{Field,Buks} --- quantum point 
 contacts (QPCs) placed nearby the quantum dots and tuned to half pinch off.
Such a charge detector is capable of sensing a change in the dot occupation 
 number by as much as a fraction of an electron~\cite{Elzerman}.
Using this method of charge control should allow one
 to concentrate in more detail
 on the transport properties of the DD.
As we show in this paper, the differential
 conductance of the DD as a function of the bias voltage
 provides valuable information about the DD parameters, 
 including the exchange constant $J$.

We begin with a Hund-Mulliken model for the DD in 
 Section~\ref{secHundMul}. 
We introduce the parameters which characterize
 the DD. 
Next, throughout the paper, we refer to these parameters 
 as phenomenological ones, 
 seeking ways to extract them from possible experiments.
In Section~\ref{sequential}, we use a master equation approach and derive
 the sequential-tunneling current $I$ at a finite bias voltage $\Delta\mu$.
Here, we find a novel feature for the DD: 
 the differential conductance $G=edI/d\Delta\mu$ as a function of $\Delta\mu$
 has a peak of {\em negative} $G$ for typical DD parameters.
We find a number of additional peaks, which all together allow one to extract 
 the exchange constant $J$ and the DD tunnel splitting $2t_0$.
In Section~\ref{cotunnelingN=1}, we consider cotunneling 
 through a Coulomb blockaded DD with one electron.
We calculate the elastic and inelastic components of the current and
 specify a temperature regime of {\em strong heating}, 
 where the asymmetry of the coupling to the
 leads, $\eta$, can be extracted from transport measurements.
The procedure is explained in detail in Section~\ref{cotunnelingN=2},
 where we consider the cotunneling through the Coulomb blockaded DD 
 with two electrons.
We show that the {\em interaction} parameter $\phi$, which gives the
 Hubbard ratio and entanglement between the electrons in the DD 
 singlet state, can be extracted from transport measurements.
Finally, in Section~\ref{together}, we consider a combined
 effect of sequential tunneling and cotunneling for the DD.
Here, again we find additional peaks or dips, which can occur in the Coulomb
 blockaded valleys for the non-linear conductance 
 below certain temperature scales, which we specify.

\section{Energy Spectrum of a Double Quantum Dot}           %
\label{secHundMul}
For definiteness we consider lateral quantum dots, which are usually
 formed by gating a two-dimensional electron gas (2DEG) under the surface  
 of a substrate. 
The 2DEG is depleted in the regions under the gates and, with an appropriate
 gate design\cite{Ciorga}, 
 one can achieve a depopulation of the dots down to 1 and 
 0 electrons per dot, avoiding pinching of the dots from the rest of the 2DEG.
The low energy sector of a DD at occupation
 number $N=N_1+N_2=1$ consists of two tunnel-split energy levels, which 
 we label by the orbital quantum number $n$, with $n=+$ standing for the 
 symmetric orbital and $n=-$ for the anti-symmetric one.
The states of the DD with one electron can then be written as
\begin{eqnarray}\label{states1}
|n,\sigma\rangle=d_{n\sigma}^{\dag}|0\rangle\;,
\end{eqnarray}
where $\sigma$ denotes the spin degeneracy of each level 
 (we neglect the Zeeman splitting), 
 $d_{n\sigma}^{\dag}$ is the electron creation operator, and $|0\rangle$ is 
 the DD state with zero electrons. 
The splitting between the two levels is given by $2t_0$, where $t_0$
 is the interdot tunneling amplitude.
We assume weak tunnel-coupling between the dots such that 
 $t_0\ll\hbar\omega_0$, where $\hbar\omega_0$ is the size-quantization
 energy of a single dot. 
Then, for the occupation number $N=N_1+N_2=2$,
 the lowest energy states, one singlet state and 3 triplet states 
 are given by
\begin{eqnarray}\label{states}
&&|S\rangle=\frac{1}{\sqrt{1+\phi^2}}(d_{+\uparrow}^{\dag}
d_{+\downarrow}^{\dag}-\phi d_{-\uparrow}^{\dag}
d_{-\downarrow}^{\dag})|0\rangle,
\nonumber\\
&&|T_+\rangle=d_{-\uparrow}^{\dag}d_{+\uparrow}^{\dag}|0\rangle,\;\;\;\;
|T_-\rangle=d_{-\downarrow}^{\dag}d_{+\downarrow}^{\dag}|0\rangle,
\nonumber\\
&&|T_0\rangle=\frac{1}{\sqrt{2}}(d_{-\uparrow}^{\dag}
d_{+\downarrow}^{\dag}+d_{-\downarrow}^{\dag}
d_{+\uparrow}^{\dag})|0\rangle.
\end{eqnarray}
The splitting between the singlet and the triplet, 
 $J=E_{|T\rangle}-E_{|S\rangle}$, plays the role of the Heisenberg exchange
 interaction for the two electron spins in the DD, 
$H_{\mbox{\rm spin}}=J{\bf S}_1\cdot{\bf S}_2$.
The {\em interaction} parameter $\phi$, 
 entering the singlet state in (\ref{states}),
 is determined by a competition between tunneling and Coulomb interaction
 in the DD, and it can be calculated\cite{DDKondo} 
 (Hund-Mulliken method) to be 
\begin{equation}
\phi=\sqrt{1+\left(\frac{4t_H}{U_H}\right)^2}-\frac{4t_H}{U_H}\;,
\label{phi1}
\end{equation}
where $t_H$ and $U_H$ are the so called extended inter-dot tunneling amplitude
 and on-site Coulomb repulsion, respectively~\cite{BLD}. 
We note that $t_H=t_0+t_C\simeq t_0$, where $t_C$ is a 
 Coulomb contribution, which vanishes with detaching the dots ($t_0\to 0$).
To illustrate the meaning of $\phi$, we also present the states (\ref{states})
 in terms of orbitals, which are mostly localized on one of the dots, 
 see Appendix~A.
The double occupancy in the singlet state is given by
\begin{equation}\label{doubocc}
D=\frac{(1-\phi)^2}{2(1+\phi^2)}.
\end{equation}
The parameter $\phi$ also determines the entanglement between the two
 electrons in the singlet state. 
While $\phi$ can be used on its own as a 
 measure of entanglement, we are presenting here a formula for the 
concurrence~\cite{Schliemann} of the singlet $|S\rangle$,
\begin{equation}\label{eta}  
c=\frac{2\phi}{1+\phi^2}.
\end{equation} 
The entanglement in the $|T_0\rangle$ state is maximal ($c=1$) 
at all values of $t_0$.

\begin{figure}\vspace{0cm}\narrowtext
{\epsfxsize=7cm
\centerline{{\epsfbox{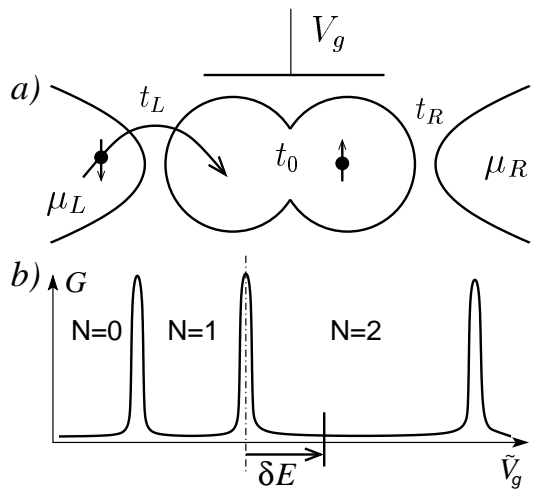}}}}
\caption{
{\em a}) 
Two coupled quantum dots,
 with the inter-dot tunneling amplitude $t_0$,
 attached to metallic leads at different chemical
 potentials $\mu_L$ and $\mu_R=\mu_L-\Delta\mu$.
The tunnel coupling between the dots and
 leads is characterized at $t_0=0$ by the tunneling
 amplitudes $t_L$ and $t_R$.
{\em b}) 
A schematic plot of the linear ($\Delta\mu\to 0$)
 conductance $G$ through the double dot (DD) as a function of the gate
 voltage $\tilde{V}_g$, showing the Coulomb blockade (CB) effect,
 with sequential-tunneling peaks and CB valleys.
The number of electrons contained in the DD, $N=N_1+N_2$,
 is fixed in the valleys between the peaks. 
The position away from the $N=1,2$ peak is measured by $\delta E$,
 which takes on negative (positive) values in the $N=1$ ($N=2$) CB valley.
}
\label{ddsetup}
\end{figure}

The energies of the considered DD states are given by
\begin{eqnarray}
\label{E_0}
&&E_{|0\rangle}=0,\\
\label{E_tun}
&&E_{|n\sigma\rangle}=\tilde{\varepsilon}-nt_0,\\
\label{E_S}
&&E_{|S\rangle}=2\tilde{\varepsilon}+U_{12}-J,\\
&&E_{|T\rangle}=2\tilde{\varepsilon}+U_{12},
\label{E_T}
\end{eqnarray}
where $\tilde{\varepsilon}\simeq\hbar\omega_0-\tilde{V}_g$ and 
 $U_{12}\simeq e^2/2C_{12}$, with $\tilde{V}_g$ being the energy shift
 due to the common gate potential $V_g$ (see Fig.~\ref{ddsetup}) 
 and $C_{12}$ the mutual capacitance between the dots.

For small dots ($\hbar\omega_0>U_H$), the next excited states are
the following two singlets
\begin{eqnarray}\label{pure}
&&|S1\rangle=\frac{1}{\sqrt{2}}(d_{-\uparrow}^{\dag}
d_{+\downarrow}^{\dag}-d_{-\downarrow}^{\dag}
d_{+\uparrow}^{\dag})|0\rangle\;,\\
&&|S2\rangle=\frac{1}{\sqrt{1+\phi^2}}(\phi d_{+\uparrow}^{\dag}
d_{+\downarrow}^{\dag}+d_{-\uparrow}^{\dag}
d_{-\downarrow}^{\dag})|0\rangle\;,
\label{out}
\end{eqnarray}
with $E_{|S1\rangle}\simeq E_{|S2\rangle}\simeq E_{|S\rangle}+U_H$. 
The states (\ref{states}) together with (\ref{pure}) and (\ref{out})
 complete the resolution of unity for two electrons in the DD
 orbitals $n=\pm$.

Finally, we note that according to the Hund-Mulliken method the
exchange constant $J$ consists of two components,
\begin{eqnarray}\label{K_0}
J&=&V_C+J_H,
\end{eqnarray}
where $V_C<0$ is responsible for a singlet-triplet transition at a finite
 magnetic field, see Ref.~\onlinecite{BLD}, and  
\begin{eqnarray}
J_H&=&\frac{1}{2}\sqrt{U_H^2+16t_H^2}-\frac{1}{2}U_H
\label{K_H}
\end{eqnarray}
 resembles the exchange constant obtained in the standard Hubbard model 
for on-site Coulomb repulsion. 
For weakly coupled quantum dots, we have 
$J_H\approx 4t_H^2/U_H$.

  \section{Sequential Tunneling through the Double Dot}                                                       %
\label{sequential}

The setup we are considering is shown in Fig.~\ref{ddsetup}{\em a}.
Each dot is tunnel-coupled to a metallic lead via a point contact,
 forming a series lead-dot-dot-lead setup [for a parallel configuration
 see Refs.~\onlinecite{DL+Suk} and~\onlinecite{Noise}].
The point contact in lateral structures has a smooth 
 (wave-guide-like) potential, providing a number of lead
 modes (channels) that can couple to the dot.
By constraining the point contact, the lead modes can be pinched
 off one by one, due to the transverse quantization in the contact region.
When the last mode in each point contact is about to be pinched off, 
 the structure shows Coulomb blockade (CB) effect at low temperatures.
In this regime, the point contacts can be treated as tunnel junctions, 
 with tunneling amplitudes $t_L$ and $t_R$, for the left and right dot, 
 respectively.
The leads are then single-channel Fermi-liquid leads, which we describe
 by the Hamiltonian
\begin{equation}\label{leads}
H_{\text{leads}}=\sum_{l}H_l=
\sum_{l}\sum_{k\sigma}\varepsilon_kc_{lk\sigma}^{\dag}
c_{lk\sigma},
\end{equation}
where $c_{lk\sigma}^{\dag}$ creates an electron with
 momentum $k$ (energy $\varepsilon_k$) and spin $\sigma$ in lead $l=L,R$.
The tunneling between the DD and the leads is described by the tunneling
 Hamiltonian
\begin{equation}\label{H_T}
H_T=\sum_{l}H_T^l=\sum_{l}\sum_{nk\sigma}(t_{ln}
c_{lk\sigma}^{\dag}d_{n\sigma}+\mbox{h.c.})\;,
\end{equation}
with the tunneling amplitudes:
\begin{equation}\label{tunamp}
t_{L,\pm}=\frac{t_L}{\sqrt{2(1\pm{\cal S})}},\;\;\;\;\;\;
t_{R,\pm}=\pm\frac{t_R}{\sqrt{2(1\pm{\cal S})}}\;.
\end{equation}
Here, ${\cal S}\sim t_0/\hbar\omega_0$ is the overlap integral between
 the two dots orbital wave functions 
 (${\cal S}=\langle\varphi_L|\varphi_R\rangle$).
Formulas (\ref{tunamp}) account for the hybridization of the DD orbitals
 at $2t_0\lesssim\hbar\omega_0$ with the accuracy of the method of molecular 
 orbitals (Hund-Mulliken method).
The tunnel-coupling to the leads broadens the DD levels $n=\pm$, introducing
 the level width $\Gamma_n=\pi\nu(|t_{Ln}|^2+|t_{Rn}|^2)$, where $\nu$ is the
 lead density of states per spin.
For our convenience, we also use the notations: 
 $\Gamma_{l}=\pi\nu|t_{l}|^2$, with $l=L,R$, and $\eta=|t_R|^2/|t_L|^2$.
Throughout the paper, we assume $\Gamma_\pm\lesssim T$, where $T$ is the
temperature.

The usual CB stability diagram\cite{kouwenhoven,KAT} 
 for a DD represents a honeycomb structure
 of increased linear conductance plotted versus $V_{g1}$, $V_{g2}$ --- 
 the gate voltages controlling each of the two dots, respectively.
We are interested in the case when the two dots are similar, and therefore
 consider the diagonal of the stability diagram, 
 $V_{g1}=V_{g2}=V_g$, in the vicinity of (1,1)-(1,0)-(0,1) triple point.
However, we will be interested in large applied bias voltages
 $\Delta\mu=\mu_L-\mu_R$, assuming
 that the bias voltage drop occurs on the structure as a whole and does 
 not shift it away from the diagonal of the stability diagram.
The bias voltage $\Delta\mu$ can be applied in different ways, e.g. 
 equally distributed between the left and right leads, or applied 
 to one of the leads alone.
To cover all possibilities, we assume the chemical potentials of the
left and right leads to be respectively equal to:
\begin{eqnarray}\label{bias}
&&\mu_L=\mu+\Delta\mu_L,\nonumber\\
&&\mu_R=\mu-\Delta\mu_R,
\end{eqnarray}
with $\Delta\mu_L+\Delta\mu_R=\Delta\mu$, and the bias-asymmetry parameter:
 $a=\Delta\mu_R/\Delta\mu_L$.
The position in the CB is controlled by the common gate voltage $V_g$.
In the linear regime ($\Delta\mu\to 0$) the conductance $G$ as a function of 
 $V_g$ shows peaks at the degeneracy points of the chemical potential in the
 DD and in the leads, see Fig.~\ref{ddsetup}{\em b}.
We focus on the sequential tunneling peak, where the number of electrons in 
 the DD fluctuates between $N=1$ and $N=2$.
The degeneracy condition is given by: $E(2)-E(1)=\mu$, where $E(N)$ is the 
 ground state energy of the DD with $N$ electrons. 
We assume the singlet (\ref{E_S}) to be the ground state for $N=2$;
 therefore, $E(2)=E_{|S\rangle}$ and $E(1)=E_{|+,\,\sigma\rangle}$.
The position in a CB valley is characterized by the addition/extraction
 energy: $E_\pm(N)=E(N\pm 1)-E(N)\mp\mu$.
The distance away from the peak (in the scale of $\tilde{V}_g\propto  V_g$) 
 is measured by
\begin{equation}\label{dE}
\delta E=E_{|+,\,\sigma\rangle}-E_{|S\rangle}+\mu.
\end{equation}
Positive (negative) values of $\delta E$ correspond to positions in the $N=2$ 
 ($N=1$) CB valley.
The size of the $N=1$ CB valley (distance between its 1st and 2nd CB peaks) 
 is given by $E_+(1)+E_-(1)=U_{12}+2t_0-J$. 
The size of the $N=2$ CB valley is of order of $U_H$ ($U_H>U_{12}$).

The sequential tunneling through the DD is described by the golden rule 
rates:
\begin{eqnarray}\label{gdrlrt1}
W_{Mm}^l=\frac{2\pi}{\hbar}\sum_{\bar{n},k\sigma}
\left|\langle M;\bar{n}|
c_{lk\sigma}^\dag H_T^l|m;\bar{n}\rangle\right|^2
\delta(\varepsilon_k-E_{Mm})\rho_{l,\bar{n}}^B,\;\;\;&&\\
\label{gdrlrt2}
W_{mM}^l=\frac{2\pi}{\hbar}\sum_{\bar{n},k\sigma}
\left|\langle m;\bar{n}|c_{lk\sigma}
H_T^l|M;\bar{n} \rangle\right|^2
\delta(\varepsilon_k+E_{Mm})\rho_{l,\bar{n}}^B,\;\;\;&&
\end{eqnarray}
where $m$ stands for one of the states (\ref{states1}),
 and $M$ for one of the states (\ref{states});
 $W_{Mm}^l$ is the probability rate for the DD transition from $m$ to $M$ 
 by exchanging an electron with the lead $l$; $E_{Mm}=E_M-E_m$ with
 $E_M$, $E_m$ being one of (\ref{E_0})$-$(\ref{E_T}). 
The averaging in (\ref{gdrlrt1}) and (\ref{gdrlrt2}) 
 is performed over the Fermi-sea states $|\bar{n}\rangle$ 
 with the grand canonical density matrix
 $\rho_l^B=Z_l^{-1}\exp(-K_l/k_BT)$, where $Z_l={\rm Tr}_l\exp(-K_l/k_BT)$, 
 $K_l=H_l-\mu_l\sum_{k\sigma}c_{lk\sigma}^{\dag}c_{lk\sigma}$,
 and $T$ is the temperature (we set $k_B=1$ in what follows).
In (\ref{gdrlrt1}) and (\ref{gdrlrt2}), we use the notations: 
 $\rho_{l,\bar{n}}^B=\langle\bar{n}|\rho_l^B|\bar{n}\rangle$ and
 $|m;\bar{n}\rangle=|m\rangle|\bar{n}\rangle$.
As a consequence of the thermal equilibrium in the leads, 
 the rates (\ref{gdrlrt1}) and (\ref{gdrlrt2})
 are related by the detailed-balance formula:
\begin{equation}\label{detbal}
W_{mM}^l=W_{Mm}^l\exp\left[(E_{Mm}-\mu_l)/T\right].
\end{equation}
We provide explicit expressions for the rates $W_{Mm}^l$ in Appendix~B.

It is convenient for the calculation to trace out
 the spin degeneracy of (\ref{states1}) and 
 of the triplets (\ref{states}).
We map the degenerate levels onto non-degenerate ones, using the
 following replacement\cite{Noise}
\begin{equation}\label{map}
\frac{1}{N_I}\sum_{i\in I\atop f\in F}W_{fi}\to W_{FI},
\end{equation}
where $N_I$ is the degeneracy of level $I$.
Thus, from here on, we deal with 4 non-degenerate levels, 
 denoted as 
 $|+\rangle$, $|-\rangle$, $|S\rangle$, $|T\rangle$. 
The transition rates between these states, for $l=L$, are:
\begin{eqnarray}
&&W_{S,+}^L=\frac{2\pi}{\hbar}\nu
\frac{|t_{L,+}|^2}{1+\phi^2}f(-\delta E-\Delta\mu_L),\nonumber\\
&&W_{+,S}^L=2\frac{2\pi}{\hbar}\nu
\frac{|t_{L,+}|^2}{1+\phi^2}f(\delta E+\Delta\mu_L),\nonumber\\
&&W_{S,-}^L=\frac{2\pi}{\hbar}\nu
\frac{\phi^2|t_{L,-}|^2}{1+\phi^2}f(-\delta E-2t_0-\Delta\mu_L),\nonumber\\
&&W_{-,S}^L=2\frac{2\pi}{\hbar}\nu
\frac{\phi^2|t_{L,-}|^2}{1+\phi^2}f(\delta E+2t_0+\Delta\mu_L),\nonumber\\
&&W_{T,+}^L=\frac{3}{2}\frac{2\pi}{\hbar}\nu
|t_{L,-}|^2f(J-\delta E-\Delta\mu_L),\nonumber\\
&&W_{+,T}^L=\frac{2\pi}{\hbar}\nu
|t_{L,-}|^2f(-J+\delta E+\Delta\mu_L),\nonumber\\
&&W_{T,-}^L=\frac{3}{2}\frac{2\pi}{\hbar}\nu
|t_{L,+}|^2f(J-\delta E-2t_0-\Delta\mu_L),\nonumber\\
&&W_{-,T}^L=\frac{2\pi}{\hbar}\nu
|t_{L,+}|^2f(-J+\delta E+2t_0+\Delta\mu_L).
\end{eqnarray}
Expressions for $l=R$ are obtained from above by substituting
 $L\to R$ and $\Delta\mu_L\to -\Delta\mu_R$. Here,
$f(E)=1/\left[1+\exp(E/T)\right]$ is the Fermi function.

\subsection{Master Equation}
\label{ME}
Assuming a large temperature $T\gg\Gamma_\pm$, but still much smaller than the
 scales of interest, we describe the DD state by the 
 probability $\rho_p$ for the DD to be in the level 
 $p\in \left\{|+\rangle,\, |-\rangle,\, |S\rangle,\, |T\rangle\right\}$.
In the stationary limit, $\rho_p$ obeys the balance equations:
\begin{eqnarray}
&&\left(W_{S,+}+W_{T,+}\right)\rho_+=
W_{+,S}\rho_S+W_{+,T}\rho_T\;,\nonumber\\
&&\left(W_{S,-}+W_{T,-}\right)\rho_-=
W_{-,S}\rho_S+W_{-,T}\rho_T\;,\nonumber\\
&&\left(W_{+,S}+W_{-,S}\right)\rho_S=
W_{S,+}\rho_++W_{S,-}\rho_-\;,\nonumber\\
&&\left(W_{+,T}+W_{-,T}\right)\rho_T=
W_{T,+}\rho_++W_{T,-}\rho_-\;,\label{msteq}
\end{eqnarray}
where we used the notation $W_{pp'}=W_{pp'}^L+W_{pp'}^R$.
Only three of the equations (\ref{msteq}) are linearly independent.
Choosing any three of them, and using the normalization condition
\begin{equation}\label{norm}
\rho_++\rho_-+\rho_S+\rho_T=1\;,
\end{equation}
one can find the solution for $\rho_p$.
However, it is convenient for the further discussion 
 to describe the non-equilibrium in the DD by the following balance ratios:
\begin{eqnarray}
\label{tu}
&&\tau=\frac{\rho_S+\rho_T}{\rho_++\rho_-}\;,\\
\label{bt}
&&\beta=\rho_T/\rho_S\;,\\
\label{gm}
&&\gamma=\rho_-/\rho_+\;,
\end{eqnarray}
which give the occupation probability $\rho_p$ of the DD states as
\begin{eqnarray}
\rho_+=\frac{1}{(1+\tau)(1+\gamma)}\;,\;\;\;
\rho_-=\frac{\gamma}{(1+\tau)(1+\gamma)}\;,&&\nonumber\\
\label{rsntr}
\rho_S=\frac{\tau}{(1+\tau)(1+\beta)}\;,\;\;\;
\rho_T=\frac{\tau\beta}{(1+\tau)(1+\beta)}\;.&&
\end{eqnarray}
Expressions for $\tau$, $\beta$, $\gamma$ are given in Appendix~C.
In the linear regime, the DD is in thermodynamic equilibrium, and
 the occupation of the states is determined by the temperature $T$.
For this regime, we find the equilibrium values:
\begin{eqnarray}\label{eq_btgmT}
&&\beta^T=3\exp(-J/T),\;\;\;\;\;\;\;\;\gamma^T=\exp(-2t_0/T),\\
\label{eq_tauT}
&&\tau^T=\frac{(1+\beta^T)\exp(\delta E/T)}{2(1+\gamma^T)}.
\end{eqnarray}
In the non-linear regime, the deviation 
 from these equilibrium values due to the applied bias 
 describe the {\em heating effect} in the DD.
To simplify our further considerations we make the following assumptions:
\begin{eqnarray}
& a)&\;\;\;\; |\delta E|>2t_0>J>0\,,\label{cond_a}\\
& b)&\;\;\;\; T\ll J,\; 2t_0-J,\; |\delta E|-2t_0\,.\label{cond_b}
\end{eqnarray}
For definiteness, we also assume a symmetric bias situation with 
 $\Delta\mu_L=\Delta\mu_R=\Delta\mu/2$.
Next, we consider the two following cases.

\subsubsection{Sequential tunneling on the $N=1$ CB valley 
 side $(\delta E<0)$.}
\label{dE<0}
As mentioned in Section~\ref{secHundMul}, 
 the $N=1$ CB valley has the size $U_{12}+2t_0-J$. 
We can estimate $U_{12}$ for a lateral structure 
 from the Hund-Mulliken method in 
 the absence of screening of Coulomb 
 interaction due to top gates.
We obtain~\cite{note1}
 $U_{12}\approx 2.7\,{\rm meV}$
 for GaAs quantum dots with $\hbar\omega_0=3\,{\rm meV}$ 
 coupled so that the distance between the dots centers is $\simeq 2a_B$, 
 with $a_B=\sqrt{\hbar/m\omega_0}$ being the Bohr radius of one dot.
For the on-site Coulomb repulsion we obtain $U_H\approx 4.5\,{\rm meV}$.
The screening from the top gates, which depends on the design of 
 the structure and on the thickness of the insulating 
 layer between the 2DEG and top gates, 
 reduces, in practice, the inter-dot Coulomb repulsion as compared 
 to the on-site one.
However, we still assume a sizable $U_{12}$ such that $U_{12}\gtrsim 2t_0$,
 and thus we can neglect the contribution from the $N=0,1$ 
 sequential-tunneling peak at 
 $\Delta\mu/2\simeq |\delta E|+J<(U_{12}+2t_0-J)/2$.

\begin{figure}\narrowtext
{\epsfxsize=9cm
\centerline{{\epsfbox{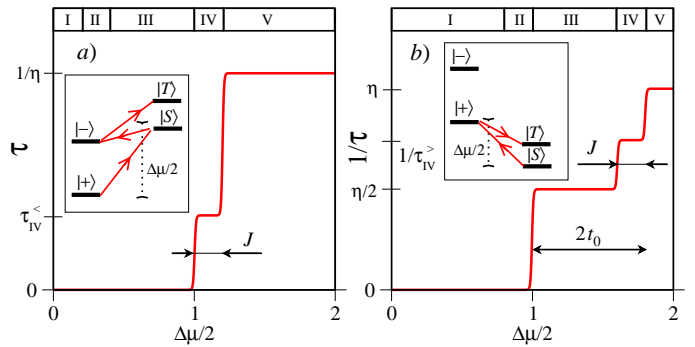}}}}
\caption{
($a$) The ratio $\tau=(\rho_S+\rho_T)/(\rho_++\rho_-)$
 plotted versus $\Delta\mu/2=\Delta\mu_{L,R}$ for $\delta E<0$.
($b$) The inverse ratio $1/\tau$ versus $\Delta\mu/2$ for $\delta E>0$.
For both figures, we choose the units of energy such that $|\delta E|=1$.
We use the following parameters: 
 $2t_0=0.8$, $J=0.2$, ${\cal S}=0.5$, and $\phi=0.4$.
The grids above the figures ($a$) and ($b$) show the division into the 
 intervals (\ref{intervals1}) and (\ref{intervals2}), {\em resp}.
The insets to the figures show the transitions which switch
 on at $\Delta\mu/2=|\delta E|$.
}
\label{seq_tau}
\end{figure}

\begin{figure}\vspace{0.cm}\narrowtext
{\epsfxsize=9cm
\centerline{{\epsfbox{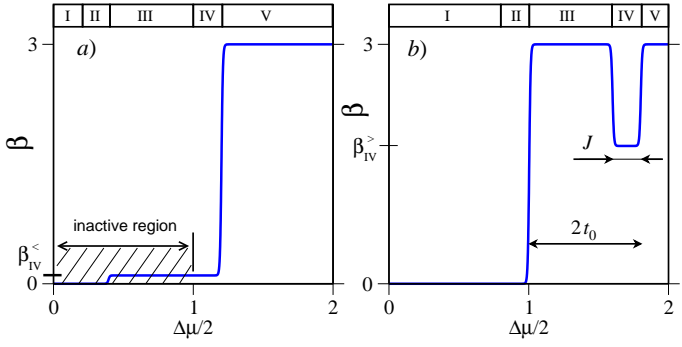}}}}
\caption{With the same parameters as in Fig.~\ref{seq_tau}, 
 the ratio $\beta=\rho_T/\rho_S$ versus $\Delta\mu/2$ for: 
 ($a$) $\delta E<0$ and ($b$) $\delta E>0$.
The shaded area in ($a$) shows the inactive region with no population of the
 $N=2$ sector, see Fig.~\ref{seq_tau}$a$.
We note that this region can be active if 
a combined effect of sequential tunneling and cotunneling is considered,
see Sec.~\ref{together}.
}
\label{seq_bt}
\end{figure}

\begin{figure}\vspace{0cm}\narrowtext
{\epsfxsize=9cm
\centerline{{\epsfbox{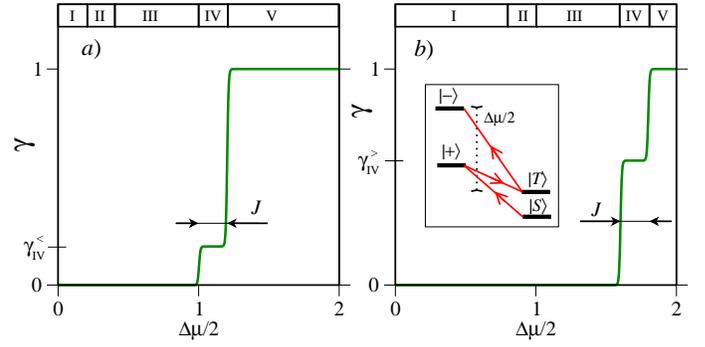}}}}
\caption{With the same parameters as in Fig.~\ref{seq_tau}, 
 the ratio $\gamma=\rho_-/\rho_+$ versus $\Delta\mu/2$ for: 
 ($a$) $\delta E<0$ and ($b$) $\delta E>0$.
The inset in ($b$) shows the sequence of transition which takes place
 at $\Delta\mu/2\geq \delta E +2t_0-J$ (compare with the inset in 
 Fig.~\ref{seq_tau}$b$).
}
\label{seq_gm}
\end{figure}

For further consideration it is convenient to divide the range of 
 the applied bias into the following intervals:
\begin{eqnarray}
&\text{I.}&\;\;\;\;\;0<\Delta\mu/2<|\delta E|-2t_0\,,\nonumber\\
&\text{II.}&\;\;\;\;\;|\delta E|-2t_0<\Delta\mu/2<|\delta E|-2t_0+J\,,\nonumber\\
&\text{III.}&\;\;\;\;\;|\delta E|-2t_0+J<\Delta\mu/2<|\delta E|\,,\nonumber\\
&\text{IV.}&\;\;\;\;\;|\delta E|<\Delta\mu/2<|\delta E|+J\,,\nonumber\\
&\text{V.}&\;\;\;\;\;|\delta E|+J<\Delta\mu/2\,.\label{intervals1}
\end{eqnarray}
At temperatures satisfying (\ref{cond_b}), 
 the solution of the master equation (\ref{msteq}) is constant 
 within each of these intervals.
We plot the quantities (\ref{tu}), (\ref{bt}) and (\ref{gm})  
 versus $\Delta\mu/2$ in Figs.~\ref{seq_tau}$a$, 
 \ref{seq_bt}$a$ and \ref{seq_gm}$a$, respectively.
The units of energy are chosen such that $|\delta E|=1$. 
Fig.~\ref{seq_tau}$a$ shows the balance between the
 DD being in the sector with $N=2$ electrons and the DD being in the
 sector with $N=1$ electrons. 
At small bias voltages (I, II and III), 
 the occupation of the $N=2$ sector is suppressed
 as $\exp\left[-(|\delta E|-\Delta\mu/2)/T\right]$, because of the Coulomb
 blockade in the $N=1$ valley.
At the bias voltage $\Delta\mu/2=|\delta E|$,
 the left lead chemical potential $\mu_L$ reaches the value of the DD
 $N=2$ chemical potential.
At this point the following sequence of transitions becomes possible:
\begin{equation}\label{cascade}
|+\rangle\rightleftarrows|S\rangle\rightleftarrows|-\rangle
\rightleftarrows|T\rangle,
\end{equation}
 which changes the population probabilities $\rho_p$ in the DD.
It is important to note the difference between the intervals IV and V.
In the interval V we get universal results: $\tau=1/\eta$, $\beta=3$, and
 $\gamma=1$. This corresponds to setting $T\to\infty$ in Eq.~(\ref{eq_btgmT}) 
 for $\beta$ and $\gamma$, and recognizing that 
 $(\rho_S+\rho_T)\sim \Gamma_L$ and $(\rho_++\rho_-)\sim \Gamma_R$, 
 which yields $\tau=\Gamma_L/\Gamma_R=1/\eta$ according to Eq.~(\ref{tu}).
In contrast, in the interval IV the plateau values of $\tau$, $\beta$ 
 and $\gamma$ depend on the DD parameters.
For example, in Fig.~\ref{seq_tau}$a$ this non-universal value of 
 $\tau$ is denoted by $\tau_{\rm IV}^<$, and we find that
\begin{equation}\label{tau4<}
1/\tau_{\rm IV}^<=\eta\left[1+
\frac{1+1/\phi^2+2\chi(2+\phi^2)}
{1+1/\phi^2+2\chi/3+8\eta/3(1+\eta)}
\right],
\end{equation}
where $\chi=\Gamma_-/\Gamma_+=(1+{\cal S})/(1-{\cal S})$.
This ``universality versus non-universality'' depends on the way
 the sequence (\ref{cascade}) is closed.
For the interval IV, only the transition $|T\rangle\to|+\rangle$
 is allowed, whereas the reverse transition is forbidden by energy 
 conservation.
For the interval V, however, the sequence is closed by 
 $|T\rangle\rightleftarrows|+\rangle$.
We can express the fact that, for the regime on the plateau V,
 the results are universal by formulating a principle of detailed
 balance.
For the rates entering the master equation (\ref{msteq}), such
 {\em a non-equilibrium detailed-balance principle} can be written as
 follows
\begin{eqnarray}
\label{dtbl1}
&&W_{S,\pm}=\frac{1}{2\eta}W_{\pm,S},\\
\label{dtbl2}
&&W_{T,\pm}=\frac{3}{2\eta}W_{\pm,T}.
\end{eqnarray}
This suffices to obtain the universal result for $\tau$, $\beta$ 
 and $\gamma$ from the master equation (\ref{msteq}).

For the Fig.~\ref{seq_bt}$a$, we find that $\beta$ changes from zero
 to a value $\beta_{\rm IV}^<$ at the border between the intervals II and III.
This is related to the fact that the transition 
 $|S\rangle\to|-\rangle\to|T\rangle$ could occur, provided there
 was a non-vanishing population in the $N=2$ sector. 
But since $\tau=0$ for this interval, see Fig.~\ref{seq_tau}$a$,
 this step-like change in $\beta$ will not be observed in physical
 quantities.
Interestingly, when the $N=2$ sector acquires non-zero 
 population at the border
 between the intervals III and IV, the ratio $\beta=\rho_T/\rho_S$ stays
 constant as a function of $\Delta\mu$ and $T$, for $T$ satisfying 
 (\ref{cond_b}).
The value of $\beta_{\rm IV}^<$ is given by
\begin{equation}\label{bt4<}
1/\beta_{\rm IV}^<=\frac{1}{3}
\left[1+(1+1/\eta)\left(\chi+\frac{3}{2}(1+1/\phi^2)\right)\right],
\end{equation}
and that of $\gamma_{\rm IV}^<$, by
\begin{equation}\label{gm4<}
1/\gamma_{\rm IV}^<=1+
\frac{3}{2}\;\frac{1+1/\phi^2+\chi(1+\phi^2)}{\chi+\eta/(1+\eta)}.
\end{equation}
The population probability of each of the levels can be
 obtained from the formulas (\ref{rsntr}). 
We plot $\rho_+$, $\rho_-$, $\rho_S$, and $\rho_T$ on Fig.~\ref{seq_rho}$a$
 for the above discussed situation.

\subsubsection{Sequential tunneling on the $N=2$ CB valley side 
 $(\delta E>0)$.}
\label{dE>0}
Here, the relevant intervals of applied bias are:
\begin{eqnarray}
&\text{I.}&\;\;\;\;\;0<\Delta\mu/2<\delta E-J\,,\nonumber\\
&\text{II.}&\;\;\;\;\;\delta E-J<\Delta\mu/2<\delta E\,,\nonumber\\
&\text{III.}&\;\;\;\;\;\delta E<\Delta\mu/2<\delta E+2t_0-J\,,\nonumber\\
&\text{IV.}&\;\;\;\;\;\delta E+2t_0-J<\Delta\mu/2<\delta E+2t_0\,,\nonumber\\
&\text{V.}&\;\;\;\;\;\delta E+2t_0<\Delta\mu/2\,,\label{intervals2}
\end{eqnarray}
and we assume $\delta E+2t_0<U_H/2$ such that the DD is not
 populated with 3 electrons while raising $\mu_L$.
We plot $1/\tau$, $\beta$ and $\gamma$ versus $\Delta\mu/2$ in 
 Figs.~\ref{seq_tau}$b$, \ref{seq_bt}$b$ and \ref{seq_gm}$b$, respectively.
At $\Delta\mu=0$ the DD is in the $N=2$ CB valley and as in the
 previous case a sizable change in the DD population occurs
 when $\Delta\mu/2=\delta E$.
At this point the chemical potential $\mu_R$ in the right lead
 is low enough such that an electron from the DD with $N=2$ can occupy
 an empty place above the Fermi sea in the right lead.
The following sequence of transition is immediately activated
\begin{equation}\label{cascade1}
|S\rangle\rightleftarrows|+\rangle\rightleftarrows|T\rangle,
\end{equation}
 which yields universal values: $\tau=2/\eta$, $\beta=3$ and $\gamma=0$,
 in the interval III. 
The corresponding principle of detailed balance is obtained if
 we disregard the level $|-\rangle$ in the master equation 
 (\ref{msteq}) and use (\ref{dtbl1}) and (\ref{dtbl2}) 
 with $\pm\to +$.
The left-to-right processes of the sequence (\ref{cascade1}) are illustrated
 in the inset to Fig.~\ref{seq_tau}$b$.
Yet two other changes in $1/\tau$ occur at  
 $\Delta\mu=\delta E+2t_0-J$ and $\Delta\mu=\delta E+2t_0$, 
 see Fig.~\ref{seq_tau}$b$.
The value of $1/\tau$ on the plateau IV in Fig.~\ref{seq_tau}$b$ 
 is given by
\begin{equation}\label{tau4>}
\tau_{\rm IV}^>=\frac{1}{\eta}\left[1+
\frac{\chi(7+3\phi^2)/4-1/2}
{3(1+1/\phi^2)/(1+\eta)+1+\chi}
\right].
\end{equation}

\begin{figure}\vspace{0cm}\narrowtext
{\epsfxsize=9cm
\centerline{{\epsfbox{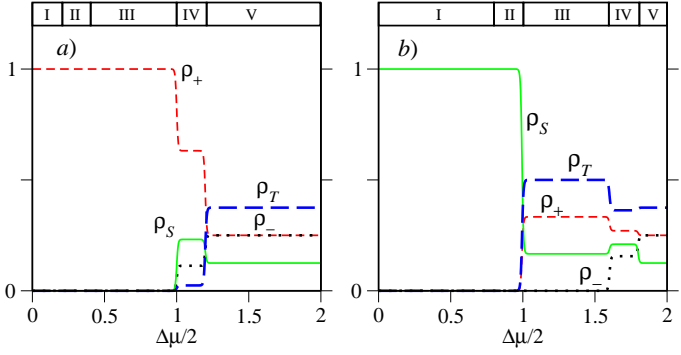}}}}
\caption{The population probabilities $\rho_p$ versus $\Delta\mu/2$
 calculated from Eqs.~(\ref{rsntr}) with the values
 of $\tau$, $\beta$ and $\gamma$ given by respectively 
 Figs.~\ref{seq_tau},~\ref{seq_bt} and~\ref{seq_gm} 
for: ($a$) $\delta E=-1$ and ($b$) $\delta E=1$.
}
\label{seq_rho}
\end{figure}

Fig.~\ref{seq_bt}$b$ shows that, in the interval IV, 
 the triplet level loses its population probability
 relative to the singlet level, with the 
 value of $\beta$ being given by
\begin{equation}\label{bt4>}
1/\beta_{\rm IV}^>=\frac{1}{3}\left[
1+\frac{2+3\chi(1+\phi^2)}{2\chi+3(1+1/\phi^2)/(1+\eta)}
\right].
\end{equation}
Just similarly to Section~\ref{dE<0}, we have here the
following sequence of transitions
\begin{equation}\label{cascade2}
|S\rangle\rightleftarrows|+\rangle\rightleftarrows|T\rangle
\rightleftarrows|-\rangle,
\end{equation}
which is closed in the interval IV by $|-\rangle\to|S\rangle$,
and in the interval V by $|-\rangle\rightleftarrows|S\rangle$.
The latter results in $\tau=1/\eta$, $\beta=3$, and $\gamma=1$,
which is identical with the universal result in Section~\ref{dE<0}.
The detailed balance for this case is also given by 
 Eqs.~(\ref{dtbl1}) and~(\ref{dtbl2}).
Finally, the non-universal value of $\gamma$ in Fig.~\ref{seq_gm} is
given by
\begin{equation}\label{gm4>}
1/\gamma_{\rm IV}^>=1+
\frac{2}{3}\;\frac{(1+\eta)(1+\chi)}{1+1/\phi^2},
\end{equation}
and we present each of the population probabilities $\rho_p$ for the 
considered case in Fig.~\ref{seq_rho}$b$.

\subsection{Sequential Tunneling Current}
\label{current}
The electron (particle) current, flowing from the DD into the lead $l$, reads
\begin{eqnarray}\label{Ileft}
I^{l}=
\left(W_{+,S}^l+W_{-,S}^l\right)\rho_S+
\left(W_{+,T}^l+W_{-,T}^l\right)\rho_T\;\;\;&&\nonumber\\
-\left(W_{S,+}^l+W_{T,+}^l\right)\rho_+-
\left(W_{S,-}^l+W_{T,-}^l\right)\rho_-.\;\;\;&&
\end{eqnarray}
In the stationary regime described by (\ref{msteq}), 
 one has $I^L=-I^R\equiv I/|e|$.
The differential conductance $G=edI/d\Delta\mu$ as
 a function of $\delta E$ and $\Delta\mu$ can be evaluated for different 
 regimes of interest.

\begin{figure}\vspace{0cm}\narrowtext
{\epsfxsize=9cm
\centerline{{\epsfbox{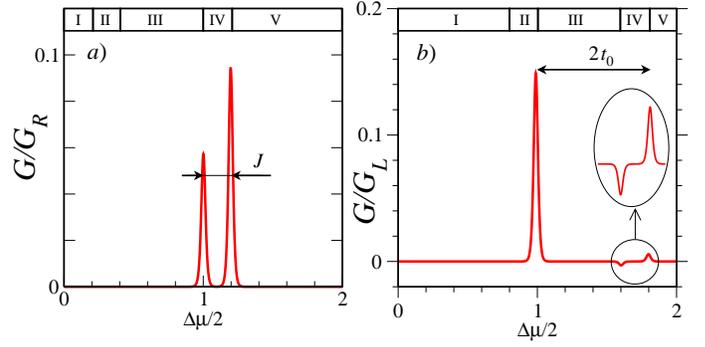}}}}
\caption{The bias dependence of the differential conductance $G$ 
 for: ({\em a}) $\delta E=-1$ and ({\em b}) $\delta E=1$. 
The ordinate axis is scaled by
  $G_l=e^2\pi\nu |t_l|^2/\hbar T$.
We use the same parameters as in Fig.~\ref{seq_tau} and with $t_L=t_R$ 
($G_L=G_R$). 
}
\label{G_dE}
\end{figure}

For the regime studied in Section~\ref{dE<0}, we can use a simplified formula
 for the current, namely
\begin{equation}\label{I_R}
I/I_0^R=\frac{2}{1+\phi^2}\left(\frac{1}{1+{\cal S}}+
\frac{\phi^2}{1-{\cal S}}\right)\rho_S+\frac{2}{1-{\cal S}^2}\rho_T\,,
\end{equation}
 where $I_0^R=|e|\Gamma_R/\hbar$.
We plot $G$ {\em vs} $\Delta\mu/2$ for $\delta E=-1$ in 
 Fig.~\ref{G_dE}$a$. 
The main peak at $\Delta\mu/2=|\delta E|=1$ acquires
 a satellite peak at $\Delta\mu/2=|\delta E|+J$, which for the given 
 parameters has a larger amplitude than the main peak. 
The origin of the satellite peak is closely related to the heating
 effects discussed in Section~\ref{dE<0}.
Eq.~(\ref{I_R}) shows that the changes in $\rho_S$ and $\rho_T$ as 
 functions of $\Delta\mu$ are directly reflected in the current $I$.

For the regime studied in Section~\ref{dE>0}, we can use a simplified formula
 for the current, namely
\begin{eqnarray}\label{I_L}
I/I_0^L&=&\left(\frac{1}{1+{\cal S}}\frac{1}{1+\phi^2}+
\frac{3/2}{1-{\cal S}}\right)\rho_+\nonumber\\
&&+\left(\frac{1}{1-{\cal S}}\frac{\phi^2}{1+\phi^2}+
\frac{3/2}{1+{\cal S}}\right)
\rho_-\,,
\end{eqnarray}
 where $I_0^L=|e|\Gamma_L/\hbar$. 
We plot $G$ {\em vs} $\Delta\mu/2$ for $\delta E=1$ in 
 Fig.~\ref{G_dE}$b$.
The main peak at $\Delta\mu/2=\delta E=1$ acquires
 two satellite peaks at $\Delta\mu/2=|\delta E|+2t_0-J$ and 
 $\Delta\mu/2=|\delta E|+2t_0$.
Interestingly, the first satellite peak has {\em negative} differential 
 conductance for the given parameter values.
Eq.~(\ref{I_L}) shows that the current $I$ reflects the changes
 in $\rho_+$ and $\rho_-$ as functions of $\Delta\mu$, discussed
 in Section~\ref{dE>0}.
The negative value of $G$ is due to the 
 decrease of $\rho_+$ when going
 from the interval III to the interval IV (see Fig.~\ref{seq_rho}$b$) 
 and different tunnel coupling to the $n=+$ and $n=-$ energy levels.
At the very origin of negative $G$ lies the Coulomb interaction in the
 DD, which allows us to consider a truncated Hilbert space, namely, 
 consisting of the states (\ref{states1}) and (\ref{states}).

\begin{figure}\vspace{0cm}\narrowtext
{\epsfxsize=9cm
\centerline{{\epsfbox{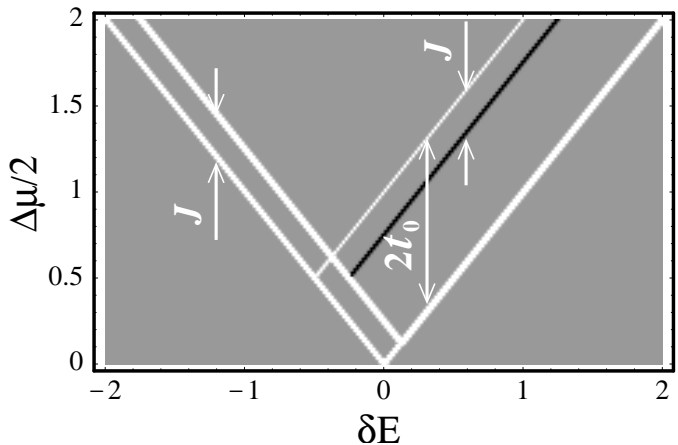}}}}
\caption{A gray scale plot of $G$ versus $\delta E$ and 
 $\Delta\mu/2=\Delta\mu_L=\Delta\mu_R$. 
The white (black) color corresponds to positive (negative) values of $G$; 
 gray stands for $G=0$.
Here, we use: $2t_0=1$, $J=0.25$, ${\cal S}=0.6$, $\phi=0.3$ and $\eta=1$.
}
\label{seq_G}
\end{figure}

Using Eqs.~(\ref{Ileft}) and (\ref{rsntr}), 
 we calculate the differential conductance
 $G$ for the whole range of variables $\delta E$ and $\Delta\mu$.
Fig.~\ref{seq_G} shows a gray-scale plot of $G$ for the
 case of symmetric biasing: $\Delta\mu_L=\Delta\mu_R=\Delta\mu/2$.
The gray color corresponds to $G=0$, the white (black) color corresponds
 to positive (negative) values of $G$.
We note that the black line on Fig.~\ref{seq_G} terminates at
 a satellite line ($\delta E=-t_0+J$, $\Delta\mu/2=t_0$), unlike the
 other two white (satellite) lines, 
 which terminate at the main sequential-tunneling peaks.
This can be attributed to the origin of the black line:
 change in the rate for {\em excited state} to {\em excited state} 
 transition, see $|T\rangle\rightarrow |-\rangle$ 
 in the inset of Fig.~\ref{seq_gm}$b$ (cf. inset of 
 Fig.~\ref{seq_tau}$b$).
For $\Delta\mu/2<t_0$, the transition $|T\rangle\rightarrow |-\rangle$
 is blocked due to energy conservation.
We note that ``excited state to excited state'' sequential-tunneling 
 satellite lines have been observed 
 experimentally for single dots\cite{Fujusawa1}.
Finally, in Fig.~\ref{seq_G1}, we present a gray-scale plot of $G$ for the
 case of asymmetric biasing: $\Delta\mu_L=\Delta\mu$ and $\Delta\mu_R=0$.

\subsection{Charge Detection via QPCs}
\label{QPC}
\begin{figure}\vspace{0cm}\narrowtext
{\epsfxsize=8cm
\centerline{{\epsfbox{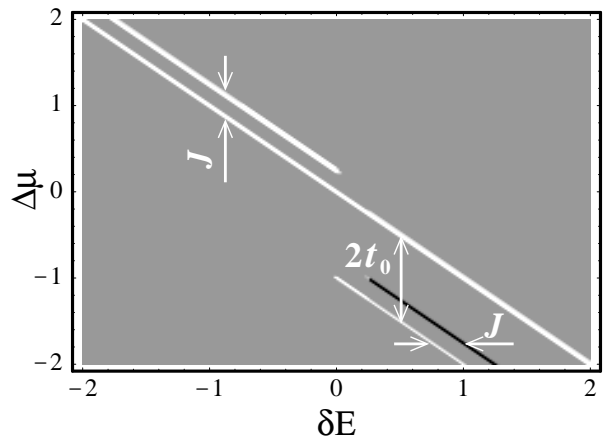}}}}
\caption{
The same as in Fig~\ref{seq_G}, but for asymmetric biasing: 
$\Delta\mu=\Delta\mu_L$, $\Delta\mu_R=0$.}
\label{seq_G1}
\end{figure}

Using quantum point contacts (QPCs) placed in the neighborhood of
 the quantum dots provides an additional channel
 of information about the DD\cite{Elzerman}.
Here, we consider the average charge on the DD, 
$\langle N\rangle=\rho_++\rho_-+2(\rho_S+\rho_T)$.
With the help of Eqs.~(\ref{tu}) and (\ref{norm}) we 
 relate $\langle N\rangle$ to the parameter $\tau$ as follows
\begin{equation}\label{<N>}
\langle N\rangle=1+\frac{\tau}{1+\tau}.
\end{equation}
At a large bias voltage, corresponding to, {\em e.g.}, the interval
 V in Section~\ref{dE<0}, see Eq.~(\ref{intervals1}), the
 DD occupation number fluctuates between 1 and 2,
 being on average  $\langle N\rangle=1+1/(1+\eta)$, for $\Delta\mu>0$.
This relation can be used to find out the asymmetry parameter $\eta$. 
In the case of symmetric coupling to the leads one has 
 $\langle N\rangle=1.5$.
In the interval IV of Section~\ref{dE<0}, the average $\langle N\rangle$
 assumes a non-universal value, determined by Eq.~(\ref{tau4<}).
This result can, in principle,  be used to extract the overlap
 integral ${\cal S}$ [or equivalently 
 $\chi=\Gamma_-/\Gamma_+=(1+{\cal S})/(1-{\cal S})$] in the case when the
 parameter $\phi$ is known, {\em e.g.} as from Section~\ref{cotunnelingN=2}.
For symmetric coupling to the leads ($\eta=1$) and in the
 limit of weakly coupled dots ($\phi\to 1$, ${\cal S}\to 0$),
 Eq.~(\ref{tau4<}) yields $\tau=1/3$, or equivalently 
 $\langle N\rangle=1.25$.
This can be distinguished from the value 
 $\langle N\rangle=1.5$ in the interval V,
 with the available sensitivity of QPCs, which is
 $\simeq 0.1\,e$, see Ref.~\onlinecite{Elzerman}.
Thus, we conclude that the satellite peaks obtained for the differential
 conductance in Section~\ref{current} (Figs.~\ref{seq_G} and~\ref{seq_G1})
 can be observed also using QPCs.~\cite{note4}
Next, we discuss the back action of the QPC on the DD.

The QPC detects the charge of the quantum dot, that it is attached to, via
 the electrostatic potential the dot induces in the QPC region.
We consider the DD and one QPC close to the right dot.
The term responsible for relaxation of the DD states
 is as follows\cite{note5}
\begin{equation}\label{Hqpc}
\delta H_{QPC}=-\frac{1}{2}\delta\varepsilon (t)\sum_{n\sigma}
d_{n,\sigma}^\dag d_{-n,\sigma},
\end{equation}
 where $\delta\varepsilon (t)$ is the fluctuating field 
 of the QPC, with $\overline{\delta\varepsilon (t)}=0$ (bar denotes
 average over the QPC degrees of freedom).
The relaxation rate is given by~\cite{Slichter}
\begin{equation}\label{Wqpc}
W_{fi}=\frac{1}{4\hbar^2}\left|\langle f|
\sum_{n\sigma}
d_{n,\sigma}^\dag d_{-n,\sigma}
|i\rangle\right|^2
J(E_{if}/\hbar),
\end{equation}
where $E_{if}=E_i-E_f$ is the transition energy.
It follows from Eq.~(\ref{Wqpc}) that the QPC can induce relaxation
 only between the states $|+,\sigma\rangle$ and $|-,\sigma\rangle$;
 all other transitions are forbidden due to the spin and charge conservation
 in the Hamiltonian (\ref{Hqpc}).
The spectral function $J(\omega)$ is defined as follows
\begin{equation}
J(\omega)=\int_{-\infty}^{\infty}\overline{\delta\varepsilon(0)
\delta\varepsilon(t)}e^{-i\omega t}dt.
\end{equation}
The QPC field, 
 $\delta\varepsilon (t)=\varepsilon (t)-\overline{\varepsilon (t)}$, 
 is proportional to the charge density at the QPC,
\begin{equation}\label{qpcfield}
\varepsilon(t)=\sum_{ll'kk'\sigma}\lambda_{ll'}
e^{i(\tilde{\mu}_l-\tilde{\mu}_{l'})t}\,
\tilde{c}^\dag_{lk\sigma}(t)\tilde{c}_{l'k'\sigma}(t),
\end{equation}
 where the indices $l,l'={\cal L,R}$ denote the two leads of the QPC 
 (not to be confused with the DD leads), $\tilde{\mu}_{l}$ is the
 chemical potential of lead $l$, 
 $\tilde{\mu}_{\cal L}=\tilde{\mu}_{\cal R}+
 \Delta\tilde{\mu}$, and
 $\tilde{c}^\dag_{lk\sigma}(t)$ creates a QPC electron.
For $\omega,\Delta\mu/\hbar<\omega_c$, we have
\begin{eqnarray}\label{qpcJ}
&&J(\omega)=4\pi\hbar\nu^2\left(\lambda_{\cal LL}^2+\lambda_{\cal RR}^2\right)
\Theta(\hbar\omega)
\;\;\;\;\;\;\;\;\;\;\;\;\;\;\;\;\;\;\;\;\;\;\;\;\;\;\;\;\;\;\;\;
\nonumber\\
&&\;\;\;\;\;\;\;\;\;\;\;
+4\pi\hbar\nu^2\lambda_{\cal LR}^2\left[\Theta(\hbar\omega+\Delta\tilde{\mu})+
\Theta(\hbar\omega-\Delta\tilde{\mu})\right],
\end{eqnarray}
 where $\Theta (E)=E/\left(1-\exp(-E/T)\right)$, $\nu$ is the density
 of states per spin in the QPC, and $\omega_c$ is the
 high-frequency cutoff (order of bandwidth).
Formula (\ref{Wqpc}) is valid for weak coupling, {\em i.e.}
 $\nu\lambda_{ll'}\ll 1$.
At $\hbar\omega, T\ll\Delta\tilde{\mu}$, Eq.~(\ref{qpcJ}) reduces to
 $J(0)=4\pi\hbar\nu^2\lambda_{\cal LR}^2\left|\Delta\tilde{\mu}\right|$,
 which formally coincides with
 $J(0)=\hbar\Gamma_d$ in the weak coupling limit, 
 where $\Gamma_d$ is the QPC decoherence rate for a single 
 dot\cite{Aleiner,Levinson}.
We summarize here by writing down the non-zero relaxation rates
 due to the QPC:
\begin{eqnarray}
\label{Wqpc+-}
&&\tilde{W}_{+-}=\frac{1}{4\hbar^2}J(2t_0/\hbar),\\
&&\tilde{W}_{-+}=\frac{1}{4\hbar^2}J(-2t_0/\hbar),
\label{Wqpc-+}
\end{eqnarray} 
 where we have already summed over the spin degeneracy, using 
 the rule (\ref{map}).
We note that, for $|\Delta\tilde{\mu}|<2t_0$, the rate $\tilde{W}_{-+}$ is
 exponentially suppressed at low temperatures.
The rates~(\ref{Wqpc+-}) and~(\ref{Wqpc-+}) describe
 relaxation of the DD due to, respectively, {\em excitation} and 
 {\em annihilation}
 of an electron-hole pair in the QPC.

Including the rate $\tilde{W}_{+-}$ into the balance equations (\ref{msteq})
 yields a correction, $\delta N$, 
 to the average occupation number on the DD,
 $\langle N'\rangle=\langle N\rangle+\delta N$, where  
 $\langle N\rangle$ is given by Eq.~(\ref{<N>}). 
In leading order in $\tilde{W}_{+-}$, the correction reads
\begin{equation}\label{dN}
\delta N=\frac{1}{(1+\tau)^2}
\frac{\gamma}{(1+\gamma)^2}
\frac{\tilde{W}_{+-}}{\tilde{W}_{0}},
\end{equation}
 where $1/\tilde{W}_{0}$ is given by Eq.~(\ref{qpcW0}).
Note that $\delta N$ is proportional to $\gamma$ and, therefore, vanishes
 in the interval III of Section~\ref{dE>0}, see Fig.~\ref{seq_gm}$b$.
In contrast,  $\delta N$ is finite in the interval V of the same section.
Thus, if the average $N$ is measured in the interval III, its value is given by
 $\langle N'\rangle=1+1/(1+\eta/2)$, see Fig.~\ref{seq_tau}$b$ and 
 Eq.~\ref{<N>};
 whereas, if the average $N$ is measured in the interval V, its value is given  by $\langle N'\rangle=1+1/(1+\eta)+\delta N$.

  \section{Cotunneling in the $N=1$ Coulomb blockade valley}     %
\label{cotunnelingN=1}

The CB valley with $N=1$ has the width $E_-(1)+E_+(1)=U_{12}+2t_0-J$.
Cotunneling dominates the conductance in the valley at $T\ll U_{12}+2t_0-J$.
In a cotunneling process, a lead electron (hole) coherently occupies the DD 
 in a state with one extra (fewer) electron and is transferred to either lead,
 leaving the DD with the same energy (elastic cotunneling) or with a different
 energy (inelastic cotunneling).
The cotunneling rate for a lead electron to go from lead $l$ to lead $l'$ 
 and the DD to go from state $m$ to state $n$ is given by the 
 golden rule rate\cite{Noise}
\begin{eqnarray}\label{rateN=1}
w_{nm}(l',l)=\frac{2\pi}{\hbar}\sum_{\bar{m}\bar{n}}
\left|\langle{\bf n}|\mbox{T}_{l'l}|{\bf m}\rangle\right|^2
\delta (E_{\bf m\,n}+\Delta\mu_{l\,l'})\rho_{\bar{m}}^B,&&\;\;\;\;
\end{eqnarray}
 where $\Delta\mu_{l\,l'}=\mu_l-\mu_{l'}$ and 
 $E_{\bf m\,n}=E_{\bf m}-E_{\bf n}$.
Here, we use the notation $E_{\bf m}=E_{m}+E_{\bar{m}}$ and 
 $|{\bf m}\rangle=|m\rangle|\bar{m}\rangle$, 
 where $E_{m}$ and $|m\rangle$ denote the DD energy (\ref{E_tun}) and
 state (\ref{states1}), {\em resp.}; furthermore, $|\bar{m}\rangle$
 is an eigenstate of $K_L+K_R$ with energy $E_{\bar{m}}$.
The averaging in (\ref{rateN=1}) is performed over the leads at thermal 
 equilibrium with the density matrix 
 $\rho^B_{\bar{m}}=\langle\bar{m}|\rho_L^B\otimes\rho_R^B|\bar{m}\rangle$.
The stationary ($\omega\to 0$) cotunneling is described by
 the effective ${\rm T}$-matrix amplitudes 
 (second order of perturbation theory)

\begin{eqnarray}
\mbox{T}_{l'l}=-\sum_{n'k'\sigma'\atop nk\sigma}t_{ln}^*t_{l'n'}
\left(\frac{d_{n\sigma}^\dag
d_{n'\sigma'}}{E_-(1)}c_{lk\sigma}c_{l'k'\sigma'}^\dag\right.\;\;\;\;\;\;\;\;\;\;\;\;\;\;\;\nonumber&&\\ 
\left.+\frac{d_{n'\sigma'}\mathbb{\char65}\,
d_{n\sigma}^\dag}{E_+(1)}c_{l'k'\sigma'}^\dag c_{lk\sigma} \right),&&
\label{Tl'l}
\end{eqnarray}
 where we assumed $J\ll E_+(1)\ll U_H$ and used 
 $\mathbb{\char65}=|S\rangle\langle S|+\sum_{i}|T_i\rangle\langle T_i|$
 to exclude virtual transitions to the states (\ref{pure}), (\ref{out}).
Formula (\ref{Tl'l}) is valid for $\Delta\mu,T\ll E_\pm(1)$.
Similarly, to Sec.~\ref{sequential} we trace out the spin degeneracy
 in (\ref{states1}), using the rule (\ref{map}). 
The cotunneling rates can then be presented as follows
\begin{equation}\label{cotratesmapped}
w_{nm}(l'l)=\frac{2\pi}{\hbar}\nu^2\Theta (E_{mn}+\Delta\mu_{ll'})
{\cal M}_{nm}^{l'l},
\end{equation}
 where $\Theta (E)=E/\left(1-\exp(-E/T)\right)$ and  ${\cal M}_{nm}^{l'l}$
 are given in Appendix~\ref{AppendixD}.
The state of the DD is described by $\rho_+=1/(1+\gamma)$ and 
 $\rho_-=\gamma/(1+\gamma)$, with 
\begin{eqnarray}
\gamma=\frac{w_{-+}}{w_{+-}}=\;\;\;\;\;\;\;\;\;\;\;\;\;\;\;\;\;\;\;\;\;\;\;\;\;\;\;\;\;\;\;\;\;\;\;\;\;\;\;\;\;\;\;\;\;\;\;\;\;\;\;\;\;\;\;\;\;\;\;\;\;\;\;\;&&\nonumber\\
\frac{\Theta(-2t_0+\Delta\mu)+\Theta(-2t_0-\Delta\mu)+
(\eta+1/\eta)\Theta(-2t_0)}{\Theta(2t_0+\Delta\mu)+\Theta(2t_0-\Delta\mu)+
(\eta+1/\eta)\Theta(2t_0)},&&\;\;\;\;\;
\label{gmheat1}
\end{eqnarray}
 where $w_{nm}=\sum_{l'l}w_{nm}(l',l)$.
We note that this result is universal and does not depend on the number
 of virtual states taken into account.
The cotunneling current is given by
\begin{equation}\label{currentI}
I=e\sum_{nm}w_{nm}^I\rho_{m},
\end{equation}
where $w_{nm}^I=w_{nm}(R,L)-w_{nm}(L,R)$.
For $T\ll 2t_0$, we define the {\em elastic} and 
 {\em inelastic} components of the current,
$I=I_{\text{el}}+I_{\text{inel}}$, as follows
\begin{eqnarray}
\label{currentIel}
&&I_{\text{el}}=ew_{++}^I,\\
\label{currentIinel}
&&I_{\text{inel}}=ew_{-+}^I\rho_++
e\left(w_{+-}^I+w_{--}^I-w_{++}^I\right)\rho_-,\;\;\;\;
\end{eqnarray}
where we used that $\sum_n\rho_n=1$.
We note that the component $I_{\text{el}}$ is a linear function of 
 $\Delta\mu$, whereas $dI_{\text{inel}}/d\Delta\mu$ 
 has a step-like $\Delta\mu$-dependence
 with the step at $\Delta\mu=2t_0$.

Next, we consider the case of a highly asymmetric coupling to the leads, 
 $(\eta+1/\eta)\gg 1$. 
For $\Delta\mu\gtrsim 2t_0$, there is a competition between two
 types of processes of inelastic cotunneling. 
One is the thermal equilibration of the DD,
 due to inelastic cotunneling into the same lead; and the second one is 
 the heating effect of the DD,
 due to inelastic cotunneling from the left lead to the right lead,
 provided $\mu_L>\mu_R$.
The strength of the former effect is proportional to $\exp(-2t_0/T)$, for 
 $T\ll 2t_0$. 
The latter effect is proportional to 
 $(T/2t_0)\eta/(1+\eta^2)$, for $\Delta\mu=2t_0$.
As a function of $T$, the crossover occurs at the energy scale
 $T_{\text{h}}=2t_0/w(\eta+1/\eta)$, where $w(x)=\ln(x\ln(x...\ln(x)))$ 
 for $x\geq e$.
At $T\gg T_{\text{h}}$, the DD is in thermal equilibrium with 
 $\gamma\equiv\rho_-/\rho_+=\exp(-2t_0/T)$.
At $T\simeq T_{\text{h}}$, the heating in the DD is governed 
 by both types of processes, and the ratio $\gamma$ depends on both $T$ and 
 $\Delta\mu$, as given by Eq.~(\ref{gmheat1}). 
At $T\ll T_{\text{h}}$, we are in the {\em strong} heating regime,
 dominated by processes of inelastic cotunneling from one lead to the other.
Here, we have
\begin{equation}\label{gammacotN=1}
\gamma=\left\{
\begin{array}{lll}
&\frac{\eta T}{2t_0(1+\eta)^2},\;\;\;\;\;\;\;\;\;
&|\Delta\mu-2t_0|\ll T,\\
&\frac{\Delta\mu-2t_0}{\Delta\mu+2t_0(1+\eta+1/\eta)},\;\;\;\;\;\;\;\;\;\;
&\Delta\mu-2t_0\gg T .
\end{array}\right.
\end{equation}
In this regime, we can extract the asymmetry parameter $\eta$, for 
 $\Delta\mu-2t_0\gg T$, in the following way
\begin{equation}\label{extract1}
\frac{4\eta}{(1+\eta)^2}=\frac{A}{1-A\frac{\Delta\mu-2t_0}{4t_0}},\;\;\;\;\;\;\;\;
A\equiv-\frac{2t_0}{\Delta G}\frac{dG}{d\Delta\mu},
\end{equation}
 where $\Delta G=G(\Delta\mu)-G(\infty)$.
The value of $G(\infty)$ is the value of $G$ at 
 $\Delta\mu-2t_0\gg 2t_0(1+\eta)^2/\eta$.
We note that Eq.~(\ref{extract1}) holds also for $\eta\simeq 1$, 
 then the energy scale $T_{\text{h}}$ coincides with $2t_0$.

  \section{Cotunneling in the $N=2$ Coulomb blockade valley}     %
\label{cotunnelingN=2}

The width of the $N=2$ CB valley is of the order of $U_H$. 
The energy scale of interest here is the exchange $J\ll U_H$.
Similarly to Section~\ref{cotunnelingN=1}, we calculate the cotunneling 
 rates with $N=2$ electrons in the DD, using the formula (\ref{rateN=1})
 with $T_{l'l}$ given by
\begin{eqnarray}
\mbox{T}_{l'l}=-\sum_{n'k'\sigma'\atop nk\sigma}t_{ln}^*t_{l'n'}
\left(\frac{d_{n\sigma}^\dag
d_{n'\sigma'}}{E_-(2)}c_{lk\sigma}c_{l'k'\sigma'}^\dag\right.\;\;\;\;\;\;\;\;\;\;\;\;\;\;\;\nonumber&&\\ 
\left.+\frac{d_{n'\sigma'}
d_{n\sigma}^\dag}{E_+(2)}c_{l'k'\sigma'}^\dag c_{lk\sigma} \right).&&
\label{TllN=2}
\end{eqnarray}
Here, we assumed $2t_0\ll E_-(2)$ and the energy splitting in the $N=3$
 sector to be much smaller than $E_+(2)$ as well as 
 $\Delta\mu,T\ll E_\pm(2)$.
The DD states $|m\rangle$, $|n\rangle$ in (\ref{rateN=1}) are now
 the singlet-triplet states (\ref{states}) and the corresponding energies
 are taken from (\ref{E_S}) and (\ref{E_T}).
After tracing out the spin degeneracy of the triplet (\ref{states}), the
 cotunneling rates are given by Eq.~(\ref{cotratesmapped}) with
 ${\cal M}_{nm}^{l'l}$ given in Appendix~\ref{AppendixE}.

\begin{figure}\vspace{0cm}\narrowtext
{\epsfxsize=9cm
\centerline{{\epsfbox{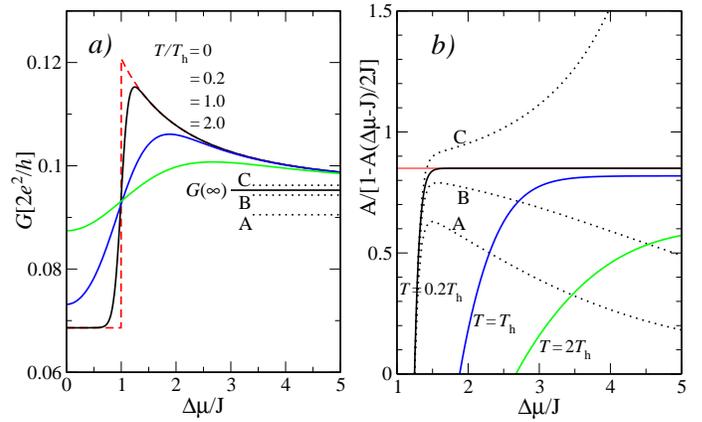}}}}
\caption{($a$)
Differential conductance $G$ {\em vs} bias
 in the cotunneling regime for the $N=2$ CB valley.
The dashed curve corresponds to the asymptotic value of $G$
 at low temperatures. The solid curves are for a finite 
 $T=(0.2,1,2)T_{\rm h}$, where $T_{\rm h}$ is the characteristic
 temperature of the strong heating regime, see text.
For the calculation we used: ${\cal S}=0.6$, $\phi=0.316$ 
 ($T_{\rm h}\simeq 0.23J$), $\eta=1$, 
 $E_-(2)=E_+(2)=E_C$, and $\Gamma_{L,R}/E_C=0.1$.
($b$)
The quantity $A/[1-A(\Delta\mu-J)/2J]$, 
 with $A\equiv -(J/\Delta G)(dG/d\Delta\mu)$, 
 where $\Delta G=G(\Delta\mu)-G(\infty)$, 
 is plotted for the three solid curves of 
 $G(\Delta\mu)$ shown in Fig.~\ref{cottunnelG}$a$.
The solid curve, corresponding to $T=0.2T_{\rm h}$, 
 saturates with good precision at the
 value of $8/(2+\kappa)=2(1-\phi)^2/(1+\phi^2)$, 
 whereas the other solid curves (not in the strong 
 heating regime) have a smaller saturation value. 
The dotted curves A, B, and C, are plotted for
 the $T=0.2T_{\rm h}$ curve
 and correspond to choosing
 the value of $G(\infty)$ as shown in Fig.~\ref{cottunnelG}$a$ 
 by dotted horizontal segments.
Only for a correct choice of $G(\infty)$ the curve has a plateau 
 and saturates at a finite value.
}
\label{cottunnelG}
\end{figure}

The following discussion is similar to the one in 
 Section~\ref{cotunnelingN=1}. 
The heating effect of the DD is described by the ratio
\begin{eqnarray}
\beta=\frac{w_{TS}}{w_{ST}}=\;\;\;\;\;\;\;\;\;\;\;\;\;\;\;\;\;\;\;\;\;\;\;\;\;\;\;\;\;\;\;\;\;\;\;\;\;\;\;\;\;\;\;\;\;\;\;\;\;\;\;\;\;\;\;\;\;&&\nonumber\\
3\frac{\Theta(-J+\Delta\mu)+\Theta(-J-\Delta\mu)+
\kappa\Theta(-J)}
{\Theta(J+\Delta\mu)+\Theta(J-\Delta\mu)+
\kappa\Theta(J)},&&\;\;\;\;\;
\label{btheat1}
\end{eqnarray}
 where $\kappa=(\eta+1/\eta)(1+\phi)^2/(1-\phi)^2$.
The DD population probabilities are given by 
 $\rho_S=1/(1+\beta)$ and  $\rho_T=\beta/(1+\beta)$. 
The formulas (\ref{currentI}), (\ref{currentIel}) and (\ref{currentIinel}) 
 apply to this case also, provided we substitute the indices 
 $+$ by $S$ and $-$ by $T$.
The differential conductance $G$, at $T\ll J$, 
 has a step-like $\Delta\mu$-dependence,
 with the step occurring at $\Delta\mu=J$.
We plot $G$ {\em vs} $\Delta\mu$ for different temperatures
 in Fig.~\ref{cottunnelG}$a$.

The strong heating regime was considered previously~\cite{GL_SST} 
 for the case $\eta=1$. 
We present the results for arbitrary $\eta$ here.
The energy scale of the {\em strong heating regime} 
 for $\kappa\gg 1$ is given by
 $T_{\text{h}}=J/w(\kappa)$, where the function $w(x)$ was introduced
 in Section~\ref{cotunnelingN=1}.
The asymptotes of $\beta$ in the strong heating regime, 
 $T\ll T_{\text{h}}$, are as follows 
\begin{equation}\label{btasympt}
\beta=\left\{
\begin{array}{lll}
&\frac{3T}{J(2+\kappa)},\;\;\;\;\;\;\;\;\;
&|\Delta\mu-J|\ll T,\\
&\frac{3(\Delta\mu-J)}{\Delta\mu+J(1+\kappa)},\;\;\;\;\;\;\;\;\;\;
&\Delta\mu-J\gg T .
\end{array}\right.
\end{equation}
For $\Delta\mu-J\gg T$, the following equality holds 
 in the strong heating regime,
\begin{equation}\label{extract2}
\frac{8}{2+\kappa}=
\frac{A}{1-A\frac{\Delta\mu-J}{2J}},\;\;\;\;\;\;\;\;
A\equiv-\frac{J}{\Delta G}\frac{dG}{d\Delta\mu},
\end{equation}
where $\Delta G=G(\Delta\mu)-G(\infty)$, with $G(\infty)$
 given by $G$ at $\Delta\mu-J\gg (2+\kappa)J$.
Eq.~ (\ref{extract2}) allows us to extract $\kappa$,
 and hence, to extract the interaction parameter $\phi$, provided
 the asymmetry parameter $\eta$ is known ({\em e.g.} from
 a similar procedure as explained in Section~\ref{cotunnelingN=1}). 
We illustrate how this procedure works in Fig.~\ref{cottunnelG}$b$.
The r.h.s. of Eq.~(\ref{extract2}) is plotted for the curves
 in Fig.~\ref{cottunnelG}$a$.
With lowering the temperature, the plateau value saturates at
 $8/(2+\kappa)$, as one enters the strong heating regime; compare
the three solid curves in Fig.~\ref{cottunnelG}$b$.
If the value of $G(\infty)$ is not known, then one can proceed in the 
following way.
Starting with a lower bound of $G(\infty)$, which can be
 {\em e.g.} $G(\infty)=0$, a set of curves is plotted
 for values of $G(\infty)$ increased each time by an offset value.
The curves can be divided into two classes: 
 (i) curves
 which have no divergence for $\Delta\mu -J>T$, have a maximum, and
 saturate at zero; 
 and (ii) curves 
 which monotonically increase, or even diverge,
 for biases $\Delta\mu -J>T$ in the available (measured) range of $\Delta\mu$.
The separatrix of these two classes has a monotonic dependence
 with a plateau at a non-zero value, and it corresponds to the asymptotic 
 value of the cotunneling conductance $G(\infty)$.
The dotted curves in Fig.~\ref{cottunnelG}$b$, denoted as A, B, and C,
 are plotted for values which might mistakenly be assigned to 
 $G(\infty)$, and are related to the curve
 at $T=0.2T_{\rm h}$.
The curves A and B belong to the class (i), and the curve C to the class (ii);
 the separatrix is the solid curve at $T=0.2T_{\rm h}$. 
The values of $G(\infty)$ taken for the curves A, B, and C, are shown in
 Fig.~\ref{cottunnelG}$a$ by dotted lines, whereas the true value of 
 $G(\infty)$ by a solid line.
Next, assuming that the measurement of $G(\Delta\mu)$ has an error bar, we
 note that the value at the maximum in curves of class (i) provides a lower
 bound for $8/(2+\kappa)$.

Finally, we note that the same physics holds true 
 for the case $J<0$ (triplet ground state),
 which can be realized by applying a magnetic field perpendicular to the 
 2DEG plane.
Here, 
 one should replace $J\to |J|$ and $\beta\to 9/\beta$ in (\ref{btasympt}),
 and
 $8/(2+\kappa)\to 8/(6+3\kappa)$ in the l.h.s. of
 (\ref{extract2}). 
Eq.~(\ref{btheat1}) remains valid for this case.

  \section{Cotunneling-assisted sequential tunneling}     %
\label{together}
In this section we consider the interplay of sequential tunneling and 
 cotunneling close to the sequential tunneling peaks at finite bias.
We find that at low temperatures, $T\ll T_0$, the heating effect due to 
 cotunneling provides population to the excited states, from which 
 a subsequent sequential tunneling can occur. 
Such a {\em cotunneling-assisted sequential tunneling} produces new
 features (peaks/dips) 
 in $G$ versus $\Delta\mu$ and $\delta E$.
The energy scale $T_0$ is given by 
 $T_0=2t_0/\ln(2t_0/\Gamma_R)$ for the $N=1$ CB side, and
 $T_0=J/\ln(J/\Gamma_L)$ for the $N=2$ CB side;
 here, we assumed symmetric biasing of the DD 
 with $\Delta\mu_L=\Delta\mu/2>0$.
For $G$ versus $\Delta\mu$ we find a peak at 
 $\Delta\mu/2=|\delta E|-2t_0>2t_0$ 
 on the $N=1$ CB valley side, as well as a peak/dip at 
 $\Delta\mu/2=|\delta E|-2t_0+J$.
On the $N=2$ CB valley side, we find a peak in $G$ versus $\Delta\mu$
 at $\Delta\mu/2=|\delta E|-J>J$.

We proceed with considering the $N=1$ CB valley side and for the sake of
 simplicity of the following expressions we assume $J\ll 2t_0\ll U_{12}$, 
 which corresponds to weakly coupled dots.
We consider
 a position in the CB valley close to the $N=1,2$ sequential tunneling peak,
 $E_-(1)\gg E_+(1)$, but still far enough to be able to apply
 a bias $\Delta\mu>2t_0$ and to have $E_+(1)-\Delta\mu_L>2t_0$. 
Since $\Delta\mu_L$ is comparable to $-\delta E=E_+(1)$,
 the cotunneling rates obtained in Section~\ref{cotunnelingN=1}
 are not valid here and
 should be modified as to account for the energy dependence of the
 tunneling density of states in the bias window.
We replace Eq.~(\ref{Tl'l}) by
\begin{eqnarray}
\mbox{T}_{l'l}=-\sum_{n'k'\sigma'\atop nk\sigma}t_{ln}^*t_{l'n'}
\frac{d_{n'\sigma'}\mathbb{\char65}\,
d_{n\sigma}^\dag}{U_+^l}c_{l'k'\sigma'}^\dag c_{lk\sigma} 
,&&
\label{TN=1close}
\end{eqnarray}
where $U_{+}^l=E(2)-E(1)-\mu_l$, and $\mathbb{\char65}$ 
 was defined below equation (\ref{Tl'l}).
Formula (\ref{TN=1close}) is valid for $T\ll U_+^l$.
Next, we calculate the cotunneling rates 
 using the golden rule expression (\ref{rateN=1}) and
 trace out the spin degeneracies according to Eq.~(\ref{map}).
We obtain the cotunneling rates close to the sequential
 tunneling peak,
\begin{equation}\label{cotratesclose}
w_{nm}(l',l)=\frac{2\pi}{\hbar}\nu^2
\left(\frac{1}{U_+^l}-\frac{1}{U_+^{l'}+E_{mn}}\right)
\tilde{\cal M}_{nm}^{l'l},
\end{equation}
for $E_{mn}+\Delta\mu_{ll'}> 0$, and 
 $w_{nm}(l',l)$  suppressed by
 $\exp\left[(E_{mn}+\Delta\mu_{ll'})/T\right]$, for 
 $E_{mn}+\Delta\mu_{ll'}<0$.
Here, we have assumed $T\ll 2t_0$ and $|E_{mn}+\Delta\mu_{ll'}|\gg T$. 
The quantities $\tilde{\cal M}_{nm}^{l'l}$ 
 are obtained from ${\cal M}_{nm}^{l'l}$
 in Appendix~\ref{AppendixD} by setting $U_-\to\infty$ and omitting
 the $1/U_+^2$ denominators.
We note that the diagonal rates $w_{nn}(l,l)$ do not enter our further
 calculation, and hence the case $E_{mn}+\Delta\mu_{ll'}=0$ refers only
 to the vicinity of $\Delta\mu=2t_0$, 
 where the value of the rate is proportional to $T$. 
For this case ($|E_{mn}+\Delta\mu_{ll'}|\ll U_+^l$), 
 one can use the expressions derived in Section~\ref{cotunnelingN=1},
 setting $E_-(1)\gg E_+(1)$.

\begin{figure}\vspace{0cm}\narrowtext
{\epsfxsize=9cm
\centerline{{\epsfbox{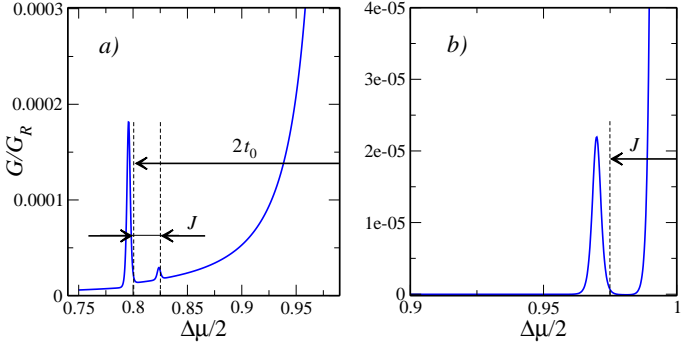}}}}
\caption{Differential conductance $G$ versus bias
 for: ({\em a}) $\delta E=-1$ and ({\em b}) $\delta E=1$.
Here, we use the following parameters: 
 $t_0=0.1$, 
 $J=0.025$,
 ${\cal S}=0.2$, 
 $\phi=0.6$,
 $\Gamma_L=\Gamma_R=0.001$,
 $T=0.001$, 
 $\Delta\mu_{L,R}=\Delta\mu/2$.
The conductance is scaled by
 $G_R=e^2\Gamma_R/\hbar T$ 
 on both plots.
}
\label{G_close}
\end{figure}

Next we solve the master equation, similar to Section~\ref{ME},
 including also the cotunneling rates.
We find that, on the $N=1$ CB valley side, 
 it suffices to take into account only the rates (\ref{cotratesclose}),
 {\em i.e.} between $|+\rangle$ and $|-\rangle$, and neglect the cotunneling
 between $|S\rangle$ and $|T\rangle$ (they give higher order correction).
From the competition of the cotunneling rate $w_{-+}$ and the sequential
 tunneling rate $W_{S,+}$ we deduce the energy scale 
 $T_0=2t_0/\ln(2t_0/\nu t_R^2)$, valid for 
 $\Delta\mu>0$~\cite{note2}.
For temperatures $T\ll T_0$, the ratio
 $\gamma=\rho_-/\rho_+$ is given by
\begin{equation}\label{gammacloseN=1}
\gamma=w_{-+}\left(w_{+-}+\frac{W_{+,S}W_{S,-}}{W_{+,S}+W_{-,S}}+
\frac{W_{+,T}W_{T,-}}{W_{+,T}+W_{-,T}}\right)^{-1}.
\end{equation}
This straightforwardly yields a non-vanishing population in the $N=2$ sector,
\begin{equation}\label{taucloseN=1}
\tau=\gamma\left(\frac{W_{S,-}}{W_{+,S}+W_{-,S}}+
\frac{W_{T,-}}{W_{+,T}+W_{-,T}}\right),
\end{equation}
 with the balance between $|T\rangle$ and $|S\rangle$ given by
\begin{equation}\label{betacloseN=1}
\beta=
\frac{W_{T,-}\left(W_{+,S}+W_{-,S}\right)}
{W_{S,-}\left(W_{+,T}+W_{-,T}\right)}.
\end{equation}
The physical interpretation of equation (\ref{gammacloseN=1})
is as follows.
In the bias range given by $U_+^L>2t_0$, the occupation of the DD is 
 determined by the cotunneling processes, with $\gamma=w_{-+}/w_{+-}$, see
 Section~\ref{cotunnelingN=1}. 
For biases $\Delta\mu>2t_0$, the population $\rho_-$ increases with 
 increasing the bias, see (\ref{gammacotN=1}),
 and can reach a value comparable to $\rho_+$.
A further increase of the bias brings us to the vicinity of the point 
 $U_+^L=2t_0$.
Here, a new channel of relaxation opens, namely 
 $|-\rangle\to|S\rangle\to|+\rangle$ (and also 
 $|-\rangle\to|T\rangle\to|+\rangle$ at $U_+^L=2t_0-J$).
Since the reverse sequence is forbidden due to energy conservation, the  
 level $|-\rangle$ is emptied efficiently and one has $\rho_+\approx 1$.
A small non-equilibrium population persists in the $N=2$ sector and 
 in the level $|-\rangle$.
This allows for a sequential tunneling current from the excited DD states.

The current through the DD consists of two parts,
\begin{equation}\label{currentcloseN=1}
I=I^{\rm seq}+I^{\rm cot},
\end{equation}
where $I^{\rm seq}$ is the sequential-tunneling
 part given by Eq.~(\ref{I_R}) with $\rho_{S,T}$
 calculated above.
The cotunneling part $I^{\rm cot}$ can be calculated with the 
 Eq.~(\ref{currentI}) and the rates (\ref{cotratesclose}). 
Both terms, $I^{\rm seq}$ and $I^{\rm cot}$ in (\ref{currentcloseN=1}),
 contribute with the same order of magnitude to the current and
 differential conductance.
In Fig.~\ref{G_close}{\em a}, we plot the bias dependence of the 
 differential conductance $G$ on the left-hand-side of the 
 sequential-tunneling peak.
The effect discussed above results in two features, namely (i)
 a peak at $\Delta\mu/2=|\delta E|-2t_0$ and (ii) a peak/dip at
 $\Delta\mu/2=|\delta E|-2t_0+J$ (both positions are given for $T=0$).
The latter peak goes into a dip with increasing the 
 inter-dot coupling\cite{note3}.
The positions of the peaks are temperature dependent.
With increasing $T$ both peaks shift to the left;
 the peak (i) shifts by $\delta\mu\propto T\ln[2t_0/\nu t_R^2(1+1/2\eta)]$ 
 and it has a width of the same order of magnitude,
 the peak (ii) shifts by $\delta\mu\sim T$ and has width $\sim T$.
This scaling behavior with varying $\Gamma_R=\pi\nu t_R^2$ comes
 about from the competition between the cotunneling rate
 $w_{+-}$ and sequential tunneling rate $W_{S,-}$ in equation 
 (\ref{gammacloseN=1}); for the peak (ii) it is not present.
We note that the peak (ii) is solely due to $dI^{\rm seq}/d\Delta\mu$,
 whereas the peak (i) is present in both $dI^{\rm seq}/d\Delta\mu$ and
 $dI^{\rm cot}/d\Delta\mu$.

Next, we proceed with considering the $N=2$ CB valley side close
 to the $N=1,2$ peak ($E_-(2)\ll E_+(2)$).
We assume $J\ll 2t_0\ll U_H$, which is usually the case for double dots, 
 and consider a position in the valley such that one can satisfy 
 $\Delta\mu>J$ and $E_-(2)-\Delta\mu_R>J$.
We follow a derivation close to the case discussed above.
Instead of Eq.~(\ref{TN=1close}) we use 
\begin{eqnarray}
\mbox{T}_{l'l}=-\sum_{n'k'\sigma'\atop nk\sigma}
t_{ln}^*t_{l'n'}
\frac{d_{n\sigma}^\dag\mathbb{\char65}\,
d_{n'\sigma'}}{U_-^{l'}}c_{lk\sigma}c_{l'k'\sigma'}^\dag
,&&
\label{TN=2close}
\end{eqnarray}
where $U_{-}^l=E(1)-E(2)+\mu_l$ and 
$\mathbb{\char65}=\sum_{\sigma}|+,\sigma\rangle\langle +,\sigma|$.
To calculate the golden rule rates we use Eq.~(\ref{cotratesclose})
 with $U_-^l\to U_+^{l'}$ and $\tilde{\cal M}_{nm}^{l'l}$ given in 
 Appendix~\ref{AppendixF}.
From solving the master equation we find results similar to 
 Eqs.~(\ref{gammacloseN=1})$-$(\ref{betacloseN=1}). 
Namely, we find that in the temperature regime $T\ll T_0$, with
 $T_0=J/\ln(J/\nu t_L^2)$ for $\Delta\mu>0$ and 
 $T_0=J/\ln(J/\nu t_R^2)$ for $\Delta\mu<0$, the cotunneling
 rate $w_{TS}$ provides non-equilibrium population to the state $|+\rangle$
 for $U_-^R\lesssim J$, where $U_-^R=E_-(2)-\Delta\mu_R$.
The ratio $\beta=\rho_T/\rho_S$ in this regime is as follows
\begin{equation}\label{betacloseN=2}
\beta=w_{TS}\left(w_{ST}+
\frac{W_{S,+}W_{+,T}}{W_{S,+}+W_{T,+}}
\right)^{-1}.
\end{equation}
The population in the $N=1$ sector is determined by
\begin{equation}\label{taucloseN=2}
\frac{1}{\tau}=\beta
\left(\frac{W_{+,T}}{W_{S,+}+W_{T,+}}\right),
\end{equation}
 and belongs to the $|+\rangle$ state, {\em i.e.} $\gamma=0$.
The current though the DD is given by Eq.~(\ref{currentcloseN=1}).
The sequential tunneling part $I^{\rm seq}$ is given by 
 Eq.~(\ref{I_L}), where one should assume 
 $\rho_-=0$, and $\rho_+=1/\tau\ll 1$ given by  Eq.~(\ref{taucloseN=2}).
The cotunneling part $I^{\rm cot}$ is given by Eq.~(\ref{currentI}), 
 where one should use occupation probabilities determined
 by Eqs.~(\ref{betacloseN=2}) and (\ref{taucloseN=2}), and cotunneling 
 rates as discussed after Eq.~(\ref{TN=2close}). 
In Fig.~\ref{G_close}{\em b}, we plot $G$ {\em vs} $\Delta\mu/2=\Delta\mu_R>0$
 on the left-hand-side of the sequential-tunneling peak.
The new feature is a peak, which occurs at $\Delta\mu/2=\delta E-J$ for $T=0$.
With increasing $T$, it moves to the left by an amount
 $\delta\mu\propto T\ln[J/\nu t_L^2(1+\eta/2)]$, and has a width of the
 same order of magnitude.
We note that the behavior of this peak is analogous to that of the
 peak (i) in the previously discussed case (see above).

To summarize, we have analyzed an effect originating from an interplay 
 of sequential tunneling and (inelastic) cotunneling, 
 and have specified the regimes when additional features 
 in transport through a DD can be observed.



                      \section{Conclusions}                      %
\label{secConclusions}
We have analyzed theoretically the transport spectroscopy of a symmetric
 DD attached to leads in series.
Our motivation was to find ways of characterizing the DD in dc 
 transport measurements.
For this, we have recast the main results of the Hund-Mulliken 
 method\cite{BLD}, introducing a description of the DD in terms of a
 set of parameters,
 $\{t_0,J,\phi,{\cal S},U_{12},U_H,\hbar\omega_0\}$,
  which can be referred to in experiments 
 as phenomenological parameters,  see Sec.~\ref{secHundMul}.
Direct access to these parameters is desirable for building spin-based
 qubits using quantum dots, see Ref.~\onlinecite{GL_SST}.
Using a master equation approach, we have described the transport 
 and the non-equilibrium probability distribution in the DD both in the
 sequential tunneling and cotunneling regimes.
We have specified a number of ``non-universal'' regimes, which reveal
 information about the DD parameters. 
We summarize our main results below.

In the sequential tunneling regime, see Sec.~\ref{sequential},
 the differential conductance at a finite bias $G(\Delta\mu)$
 has satellite peaks of sequential tunneling with respect to the
 main peaks of the CB diamond, 
 see Figs.~\ref{seq_G},~\ref{seq_G1}, and~\ref{G_dE}.
The exchange coupling constant $J$ appears as a peak separation 
 in $G(\Delta\mu)$ both on the $N=1$ and $N=2$ CB sides.
The tunnel splitting $2t_0$ can be extracted from  
 $G(\Delta\mu)$ on the $N=2$ CB side, see Fig.~\ref{G_dE}.
We note that if the two dots are detuned by some energy
 $\Delta E\lesssim 2t_0$, then one can replace  
 $2t_0\to\sqrt{4t_0^2+\Delta E^2}$ for the
 transport spectroscopy of the DD.

In the cotunneling regime, see  Secs.~\ref{cotunnelingN=1} 
 and~\ref{cotunnelingN=2}, the exchange coupling constant $J$ can be
 observed on the $N=2$ CB side as the bias value at which the inelastic 
 cotunneling turns on. 
A step in $G(\Delta\mu)$ occurs at $|\Delta\mu|=|J|$.
Similarly the tunnel splitting $2t_0$ can be extracted from the $N=1$ CB side.
In Sec.~\ref{cotunnelingN=1} we have introduced a {\em strong heating} regime
 where the bias dependence of $G(\Delta\mu)$ allows one to extract
 the asymmetry parameter $\eta=|t_R|^2/|t_L|^2$ on the $N=1$ CB side.
An additional relation involving $\eta$ and $\phi$ can be obtained on the 
 $N=2$ CB side, see Sec.~\ref{cotunnelingN=1}.

As an alternative to transport measurements we have considered the
 use of a charge detector (QPC) close to one of the dots, see 
 Sec.~\ref{QPC}.
We found that measuring the average charge on the DD allows
 one to extract $J$ and $2t_0$, similarly to the conductance measurements 
 in the sequential tunneling regime.
Moreover, additional relations between the DD parameters can be obtained
 with sensitive QPC measurements. 
We have also considered the back action of the QPC onto the DD and found
 that the QPC induces relaxation of the DD states with 1 electron.
We accounted for this relaxation in the master equations in Sec.~\ref{QPC}.

Finally, we have analyzed a combined mechanism of sequential tunneling 
and cotunneling, see Sec.~\ref{together}. 
We found that inelastic cotunneling can provide non-equilibrium population
 probability to the excited DD states, which can then allow for sequential 
 tunneling via an excited DD state.
Accounting for this effect results in additional satellite peaks in 
 $G(\Delta\mu)$ at low temperatures.

  \appendix\section{}                      %
\label{AppendixA}
Using the definition
\begin{eqnarray}\label{dntilde}
d_{n\sigma}=\frac{1}{\sqrt{2}}\left(\tilde{d}_{L\sigma}+
n\tilde{d}_{R\sigma}\right),
\end{eqnarray}
we rewrite (\ref{states}) in terms of the new operators 
$\tilde{d}_{l\sigma}$ as follows
\begin{eqnarray}\label{statesLR}
&&|S\rangle=\frac{1}{2\sqrt{1+\phi^2}}\left\{
(1+\phi)(
\tilde{d}_{L\uparrow}^{\dag}
\tilde{d}_{R\downarrow}^{\dag}-
\tilde{d}_{L\downarrow}^{\dag}
\tilde{d}_{R\uparrow}^{\dag})+\right.\nonumber\\
&&\;\;\;\;\;\;\;\;\;\;\;\;\;\;\;\;\;\;\;\;\left.(1-\phi)(
\tilde{d}_{L\uparrow}^{\dag}
\tilde{d}_{L\downarrow}^{\dag}+
\tilde{d}_{R\uparrow}^{\dag}
\tilde{d}_{R\downarrow}^{\dag})\right\}
|0\rangle\;,
\nonumber\\
&&|T_+\rangle=\tilde{d}_{L\uparrow}^{\dag}
\tilde{d}_{R\uparrow}^{\dag}|0\rangle\;,\;\;\;\;
|T_-\rangle=\tilde{d}_{L\downarrow}^{\dag}
\tilde{d}_{R\downarrow}^{\dag}|0\rangle\;,\\
&&|T_0\rangle=\frac{1}{\sqrt{2}}(
\tilde{d}_{L\uparrow}^{\dag}
\tilde{d}_{R\downarrow}^{\dag}+
\tilde{d}_{L\downarrow}^{\dag}
\tilde{d}_{R\uparrow}^{\dag})|0\rangle\nonumber\;.
\end{eqnarray}
Similarly, for (\ref{pure}) and (\ref{out}) we have
\begin{eqnarray}\label{pureLR}
&&|S1\rangle=\frac{1}{\sqrt{2}}(
\tilde{d}_{L\uparrow}^{\dag}
\tilde{d}_{L\downarrow}^{\dag}-
\tilde{d}_{R\uparrow}^{\dag}
\tilde{d}_{R\downarrow}^{\dag}
)|0\rangle\;,\\
&&|S2\rangle=
\frac{1}{2\sqrt{1+\phi^2}}\left\{
(\phi-1)(
\tilde{d}_{L\uparrow}^{\dag}
\tilde{d}_{R\downarrow}^{\dag}-
\tilde{d}_{L\downarrow}^{\dag}
\tilde{d}_{R\uparrow}^{\dag})+\right.\nonumber\\
&&\;\;\;\;\;\;\;\;\;\;\;\;\;\;\;\;\;\;\;\;\left.(1+\phi)(
\tilde{d}_{L\uparrow}^{\dag}
\tilde{d}_{L\downarrow}^{\dag}+
\tilde{d}_{R\uparrow}^{\dag}
\tilde{d}_{R\downarrow}^{\dag})\right\}
|0\rangle\;.
\label{outLR}
\end{eqnarray}

            \section{}                       %
\label{AppendixB}
The sequential tunneling rates $W_{Mm}^l$ calculated
 according to the formula (\ref{gdrlrt1}) are: 
\begin{eqnarray}
&&W_{\langle S|+,\,\uparrow\rangle}^l=\frac{2\pi}{\hbar}\nu
\frac{|t_{l,\,+}|^2}{1+\phi^2}f(E_{|S\rangle}-E_{|+,\,\uparrow\rangle}-
\mu_l)\;,\nonumber\\
&&W_{\langle S|+,\,\downarrow\rangle}^l=\frac{2\pi}{\hbar}\nu
\frac{|t_{l,\,+}|^2}{1+\phi^2}f(E_{|S\rangle}-E_{|+,\,\downarrow\rangle}-
\mu_l)\;,\nonumber\\
&&W_{\langle S|-,\,\uparrow\rangle}^l=\frac{2\pi}{\hbar}\nu
\frac{\phi^2|t_{l,\,-}|^2}{1+\phi^2}f(E_{|S\rangle}-E_{|-,\,\uparrow\rangle}-
\mu_l)\;,\nonumber\\
&&W_{\langle S|-,\,\downarrow\rangle}^l=\frac{2\pi}{\hbar}\nu
\frac{\phi^2|t_{l,\,-}|^2}{1+\phi^2}f(E_{|S\rangle}-E_{|-,\,\downarrow\rangle}-
\mu_l)\;,\nonumber\\
&&W_{\langle T_+|+,\,\uparrow\rangle}^l=\frac{2\pi}{\hbar}\nu
|t_{l,\,-}|^2f(E_{|T_+\rangle}-E_{|+,\,\uparrow\rangle}-
\mu_l)\;,\nonumber\\
&&W_{\langle T_+|+,\,\downarrow\rangle}^l=0\;,\nonumber\\
&&W_{\langle T_+|-,\,\uparrow\rangle}^l=\frac{2\pi}{\hbar}\nu
|t_{l,\,+}|^2f(E_{|T_+\rangle}-E_{|-,\,\uparrow\rangle}-
\mu_l)\;,\nonumber\\
&&W_{\langle T_+|-,\,\downarrow\rangle}^l=0\;,\nonumber\\
&&W_{\langle T_-|+,\,\uparrow\rangle}^l=0\;,\nonumber\\
&&W_{\langle T_-|+,\,\downarrow\rangle}^l=\frac{2\pi}{\hbar}\nu
|t_{l,\,-}|^2f(E_{|T_-\rangle}-E_{|+,\,\downarrow\rangle}-
\mu_l)\;,\nonumber\\
&&W_{\langle T_-|-,\,\uparrow\rangle}^l=0\;,\nonumber\\
&&W_{\langle T_-|-,\,\downarrow\rangle}^l=\frac{2\pi}{\hbar}\nu
|t_{l,\,+}|^2f(E_{|T_-\rangle}-E_{|-,\,\downarrow\rangle}-
\mu_l)\;,\nonumber\\
&&W_{\langle T_0|+,\,\uparrow\rangle}^l=\frac{2\pi}{\hbar}\nu
\frac{|t_{l,\,-}|^2}{2}f(E_{|T_0\rangle}-E_{|+,\,\uparrow\rangle}-
\mu_l)\;,\nonumber\\
&&W_{\langle T_0|+,\,\downarrow\rangle}^l=\frac{2\pi}{\hbar}\nu
\frac{|t_{l,\,-}|^2}{2}f(E_{|T_0\rangle}-E_{|+,\,\downarrow\rangle}-
\mu_l)\;,\nonumber\\
&&W_{\langle T_0|-,\,\uparrow\rangle}^l=\frac{2\pi}{\hbar}\nu
\frac{|t_{L+}|^2}{2}f(E_{|T_0\rangle}-E_{|-,\,\uparrow\rangle}-
\mu_l)\;,\nonumber\\
&&W_{\langle T_0|-,\,\downarrow\rangle}^l=\frac{2\pi}{\hbar}\nu
\frac{|t_{L+}|^2}{2}f(E_{|T_0\rangle}-E_{|-,\,\downarrow\rangle}-
\mu_l)\;,\nonumber
\end{eqnarray}
where $f(E)=1/\left[1+\exp(E/k_BT)\right]$.
The rates for the reverse transitions can be obtained from
the above formulas by replacing $f(E)\to 1-f(E)$, satisfying (\ref{detbal}).

             \section{}                     %
\label{AppendixC}
Solving (\ref{msteq}) and (\ref{norm}), we find for 
 (\ref{tu}), (\ref{bt}), and (\ref{gm}) the following expressions:
\widetext
\begin{eqnarray}
&&\tau=\frac{\left(W_{T,-}W_{-,S}+W_{-,T}W_{S,-}\right)
\left(W_{S,+}+W_{T,+}\right)+
\left(W_{S,+}W_{+,T}+W_{T,+}W_{+,S}\right)
\left(W_{T,-}+W_{S,-}\right)}
{\left(W_{+,S}W_{S,-}+W_{-,S}W_{S,+}\right)
\left(W_{+,T}+W_{-,T}\right)+
\left(W_{+,T}W_{T,-}+W_{-,T}W_{T,+}\right)
\left(W_{+,S}+W_{-,S}\right)}\;,\\
&&\beta=\frac{W_{T,+}W_{+,S}
\left(W_{S,-}+W_{T,-}\right)+
W_{T,-}W_{-,S}
\left(W_{S,+}+W_{T,+}\right)}
{W_{+,T}W_{S,+}
\left(W_{S,-}+W_{T,-}\right)+
W_{-,T}W_{S,-}
\left(W_{S,+}+W_{T,+}\right)}\;,\\
&&\gamma=\frac{W_{-,S}W_{S,+}
\left(W_{+,T}+W_{-,T}\right)+
W_{-,T}W_{T,+}
\left(W_{+,S}+W_{-,S}\right)}
{W_{+,S}W_{S,-}
\left(W_{+,T}+W_{-,T}\right)+
W_{+,T}W_{T,-}
\left(W_{+,S}+W_{-,S}\right)}\;.
\end{eqnarray}

\begin{equation}\label{qpcW0}
\tilde{W}_{0}^{-1}=\frac{(W_{S+}-W_{S-})(W_{+T}+W_{-T})+(W_{T+}-W_{T-})
(W_{+S}+W_{-S})}
{W_{+S}W_{S-}(W_{+T}+W_{-T})+W_{+T}W_{T-}
(W_{+S}+W_{-S})}.
\end{equation}
             \section{}                     %
\label{AppendixD}
\begin{eqnarray}
&&{\cal M}_{++}^{RL}=\left(\frac{1}{U_-^2}+\frac{1}{1+\phi^2}
\frac{1}{U_-U_+}+\frac{1}{(1+\phi^2)^2}\frac{1}{U_+^2}\right)
2\left|t_{L+}t_{R+}\right|^2+\frac{3\left|t_{L-}t_{R-}\right|^2}{2U_+^2}-
\frac{3}{U_-U_+}\Re\left(t_{R+}^*t_{R-}t_{L+}t_{L-}^*\right),\;\;\;\;\\
&&{\cal M}_{--}^{RL}=\left(\frac{1}{U_-^2}+\frac{\phi^2}{1+\phi^2}
\frac{1}{U_-U_+}+\frac{\phi^4}{(1+\phi^2)^2}\frac{1}{U_+^2}\right)
2\left|t_{L-}t_{R-}\right|^2+\frac{3\left|t_{L+}t_{R+}\right|^2}{2U_+^2}-
\frac{3}{U_-U_+}\Re\left(t_{R+}^*t_{R-}t_{L+}t_{L-}^*\right),\;\;\;\;\\
&&{\cal M}_{-+}^{RL}=
\left(\frac{2}{U_-^2}+\frac{3}{U_-U_+}+\frac{3}{2U_+^2}\right)
\left|t_{L-}t_{R+}\right|^2+
\frac{2\phi^2\left|t_{L+}t_{R-}\right|^2}{(1+\phi^2)^2U_+^2}-
\frac{2\phi\Re\left(t_{R+}^*t_{R-}t_{L+}t_{L-}^*\right)}{(1+\phi^2)U_-U_+},\;\;\;\;\label{M1pm}
\end{eqnarray}
where $U_-=E_-(1)$ and $U_+=E_+(1)$, and $\Re$ is real part.
For ${\cal M}_{+-}^{RL}$, change $t_{l+}\rightleftarrows t_{l-}$ 
in Eq.~(\ref{M1pm}); for ${\cal M}_{nm}^{RR}$, set $L\to R$.

             \section{}                     %
\label{AppendixE}
\begin{eqnarray}
&&{\cal M}_{SS}^{RL}=2\left|\left(\frac{\phi^2}{U_+}
-\frac{1}{U_-}\right)\frac{t_{L+}^*t_{R+}}{1+\phi^2}+
\left(\frac{1}{U_+}-\frac{\phi^2}{U_-}\right)\frac{t_{L-}^*t_{R-}}{1+\phi^2}
\right|^2,\;\;\;\;\\
&&{\cal M}_{TT}^{RL}=
\left[\left(\frac{U_++U_-}{U_+U_-}\right)^2+\frac{1}{2}\left(\frac{U_+-U_-}{U_+U_-}\right)^2\right]
\left|t_{L+}^*t_{R+}+t_{L-}^*t_{R-}\right|^2,
\;\;\;\;\\
&&{\cal M}_{ST}^{RL}=
\left(\frac{U_++U_-}{U_+U_-}\right)^2
\frac{\left|t_{L+}^*t_{R-}+\phi t_{L-}^*t_{R+}\right|^2}{1+\phi^2}=
\frac{1}{3}{\cal M}_{TS}^{LR},
\;\;\;\;\label{M2ST}
\end{eqnarray}
where $U_-=E_-(2)$ and $U_+=E_+(2)$.

\endwidetext

             \section{}                     %
\label{AppendixF}
\begin{eqnarray}
&&\tilde{\cal M}_{SS}^{RL}=
\frac{2\left|t_{L+}t_{R+}\right|^2}
{\left(1+\phi^2\right)^2},\;\;\;\;\\
&&\tilde{\cal M}_{TT}^{RL}=
\frac{3}{2}\left|t_{L-}t_{R-}\right|^2,\;\;\;\;\\
&&\tilde{\cal M}_{ST}^{RL}=
\frac{\left|t_{L+}t_{R-}\right|^2}{1+\phi^2}=
\frac{1}{3}\tilde{\cal M}_{TS}^{LR}.
\end{eqnarray}

                        \acknowledgements                        %
\addcontentsline{toc}{section}{Acknowledgments}
We thank L.~Vandersypen, J.~Elzerman, R.~Hanson, L.~Kouwenhoven, and 
 E.~Sukhorukov for discussions. 
We acknowledge support from the Swiss NSF, NCCR Nanoscience Basel, 
 DARPA, and ARO.



\begin{thebibliography}{}
\bibitem{note5}
This term, Eq.~(\ref{Hqpc}), can be obtained from
 $H=\delta\varepsilon (t)
 \sum_\sigma \tilde{d}^\dag_{R,\sigma}\tilde{d}_{R,\sigma}$,
 with the help of Eq.~(\ref{dntilde}).
More rigorously, the left dot counterpart should also be included, however,
it merely redefines $\lambda_{ll'}$ of 
Eq.~(\ref{qpcfield}).
\bibitem{note3}
For the parameters used in Fig.~\ref{G_close}, $G$ 
 has two peaks (and not a peak and a dip) 
 on the $N=2$ CB valley side for $\Delta\mu/2>\delta E$.
\end{thebibliography}
\end{document}